\documentclass[11pt,preprint]{elsarticle}
\usepackage[english,activeacute]{babel}
\usepackage{graphicx}
\usepackage{amsmath}
\usepackage{dcolumn}
\usepackage{subfigure}
\usepackage{caption}
\usepackage{amsmath,amssymb,accents}
\usepackage{units}
\usepackage{tabularx}
\usepackage{lineno}
\usepackage{amssymb}

\usepackage{hyperref}
\hypersetup{
    colorlinks=true,
    linkcolor=blue,
    filecolor=blue,      
    urlcolor=blue
    }
\makeindex

\begin{document}

\begin{frontmatter}

\title{
Uncertainties in the Estimation of Air Shower Observables \\ 
from Monte Carlo Simulation of Radio Emission 
}

\author[inst1,inst2,inst3]{Carlo~S. Cruz Sanchez}
\author[inst1,inst2,inst3]{Patricia~M. Hansen}

\author[inst1,inst2]{Matias Tueros}
\author[inst4]{Jaime Alvarez-Mu\~niz}
\author[inst5]{\\Diego~G. Melo}

\address[inst1]{{IFLP (CCT La Plata-CONICET-UNLP)},
{Diagonal 113 y 63}, 
{La Plata},
{1900}, 
{Argentina}}

\address[inst2]{{Departamento de Física, Fac. de Cs. Exactas},
{Univ. Nac. de La Plata, C. C. 67}, 
{La Plata},
{1900}, 
{Argentina}}

\address[inst3]{{Departamento de Ciencias Básicas, Facultad de Ingeniería},
{Univ. Nac. de La Plata,Calle 115 y 49}, 
{La Plata},
{1900}, 
{Argentina}}

\address[inst4]{{Instituto Galego de Física de altas Enerxias (IGFAE), Univ. de Santiago de Compostela},
{Santiago de Compostela},
{15782}, 
{Spain}}

\address[inst5]{{ITeDA (CNEA - CONICET - UNSAM), CAC - CNEA; },
{Av. Gral Paz 1499}, 
{San Martin - Buenos Aires},
{1650}, 
{Argentina}}


\begin{abstract}
The detection of extensive air showers (EAS) induced by cosmic rays via radio signals has undergone significant advancements in the last two decades. Numerous ultra-high energy cosmic ray experiments routinely capture radio pulses in the MHz to GHz frequency range emitted by EAS. The Monte Carlo simulation of these radio pulses is crucial to enable an accurate reconstruction of the primary cosmic ray energy and to infer the composition of the primary particles.  
In this work, a comprehensive comparison of the predicted electric field in EAS simulated with CoREAS and ZHAireS was conducted to estimate the systematic uncertainties arising from the use of different simulation packages in the determination of two key shower observables namely, the electromagnetic energy of the EAS and the depth of maximum development ($X_{\rm max}$). 
For this comparison, input parameters and settings as similar as possible were used in both simulations, along with the same realistic atmospheric refractive index depending on altitude, which is crucial for the prediction of radio emission properties of EAS. In addition, simulated EAS with very similar values of depth of maximum development were selected. Good agreement was found between CoREAS and ZHAireS, with discrepancies in the dominant electric field components generally remaining below 10\% across the frequency range of a few MHz to hundreds of MHz, relevant for most radio detection experiments, translating into uncertainties in the determination of energy 
below $5\%$ and $\simeq 10\,\mathrm{g/cm^2}$ in $X_{\rm max}$. Our work underscores the need for further studies to clarify their origin and impact on $X_{\max}$ inference in composition analyses.
\end{abstract}


\end{frontmatter}


\section{Introduction}
\label{S:Introduction}

Ultra-high-energy cosmic rays (UHECRs) with energies in and above the EeV range initiate extensive air showers (EAS) upon interacting with the upper layers of the Earth's atmosphere. Understanding the properties of EAS is essential to reveal the nature and origin of UHECRs \cite{Coleman:2022abf}. Various detection techniques are used for this purpose. Arrays of particle detectors sample the shower front on the ground or underground, allowing measurements of particle densities and arrival times. Fluorescence detectors capture the fluorescence light resulting from the excitations of air molecules in the atmosphere, offering information about the longitudinal shower development. Additionally, arrays of radio antennas on the ground measure the electric field emitted in the MHz\,--\,GHz frequency range by the electromagnetic particles in the shower, offering a wealth of information \cite{Schroder:2016hrv,Huege:2017khw}. This so-called radio technique has proven to be an invaluable method for EAS detection due to its almost 100\% duty cycle and the negligible amount of atmospheric absorption that radio signals experience. In addition, it has been shown that using radio detection allows an accurate determination of shower energy \cite{Tunka-Rex:2015zsa,PierreAuger:2016vya,PierreAuger:2015hbf} and the depth in the atmosphere ($X_\text{max}$) at which the shower reaches its maximum development \cite{Tunka-Rex:2015zsa,Bezyazeekov:2018yjw,Corstanje:2021kik,PierreAuger:2023lkx,PierreAuger:2023rgk,Corstanje:2025wbc}.

Radio signals in EAS primarily originate from the charged electromagnetic component. The motion of charged particles is akin to time-varying electric currents, resulting in coherent radiation for wavelengths greater than the typical dimensions of the shower \cite{Askaryan:1965}.
Since air has a refractive index $n>1$, particles in the shower can move with a speed larger than the local speed of light, inducing Cherenkov-like or relativistic effects in the emitted radiation \cite{deVries:2011pa}.  Two main effects are responsible for the bulk of the emission in EAS in the MHz frequency range. The \textit{geomagnetic effect} \cite{Kahn-Lerche:1966} involves a drift current resulting from charge separation within the magnetic field of the Earth $\vec B$, causing a polarization pattern approximately parallel to $-\hat v \times \vec{B}$ (i.e., along the direction of the Lorentz force), where $\hat v$ denotes the direction of the shower axis. The strength of the geomagnetically-induced field is directly proportional to the magnetic field magnitude $\vert \vec B \vert$ and the sine of the ``geomagnetic angle'', the angle between $\vec v$ and $\vec B$. 
The \textit{Askaryan effect} \cite{Askaryan:1962} arises from an excess of electrons compared to positrons within the shower, induced by particle physics processes such as Compton, M\"oller, and Bhabha scattering, as well as electron-positron annihilation \cite{Zas:1991jv}. The induced polarization is approximately in the direction $\hat u \times (\hat u \times \hat v)$, where $\hat u$ is a unit vector from the emission point to the observer.

The interpretation of data collected at experiments employing the radio technique requires accurate simulations of the EAS and the radio pulses produced.
The two main Monte Carlo simulation programs used for this purpose are CoREAS \cite{Huege:2013vt} and ZHAireS \cite{Alvarez_Muniz_2011ref}. In both, electrons and positrons are tracked using various approximations, and the resulting electric field is obtained from similar algorithms that are derived from first principles solving Maxwell's equations (see for instance \cite{Alvarez-Muniz:2022uey} for a review and references therein). The radio emission of the particle cascade arises naturally from these calculations, that do not assume any {\it a priori} emission mechanism or incorporate any free parameters.

Simulations conducted using ZHAireS and CoREAS have demonstrated consistent results among themselves~\cite{Huege:2013vu} and in relation to experimental data~\cite{Nelles:2014dja,PierreAuger:2016vya,PierreAuger:2018pmw,Bechtol:2021tyd}. Further comparisons between CoREAS and ZHAireS have been recently made within the CORSIKA8 code \cite{CORSIKA8-radio}. However, comparisons between the two programs have focused on specific shower geometries~\cite{Huege:2013vu}, different frequency bands~\cite{Gottowik:2017wio, Rauch:2013nxd}, or have only studied the effect on the energy emitted in radio waves~\cite{Gottowik:2017wio}. 
A more detailed comparison that spans a broader range of primary particle types, shower geometries, and frequencies - and includes both radio energy measurements and the reconstruction of the depth of shower maximum - has yet to be fully conducted. This works seeks to contribute to these aspects.

Complementing the work of \cite{Gottowik:2017wio}, in our study we provide a comparison of CoREAS and ZHAireS predictions on more fundamental observables, focusing on the lateral distribution of the different components of the electric field at ground level and their corresponding frequency spectra (Section \ref{S:Results}). Rather than simulating a large number of showers and averaging results, as done in \cite{Gottowik:2017wio}, we simulate a smaller set of showers, all with identical energy, zenith and azimuthal angles, geomagnetic angles, and similar depths of maximum. This approach avoids the need to correct for geomagnetic and Askaryan contributions to the emission at different geomagnetic angles and varying atmospheric densities at $X_\text{max}$, allowing for a more direct comparison between the two simulation codes. 

A key novelty in our study is the use of a new version of ZHAireS which incorporates the refractive index model of the atmosphere based on 
the application of the Gladstone-Dale relation between atmospheric density and refractive index \cite{Gladstone-Dale} (Section \ref{S:Simulations}). In this way the same refractive index model is used in CoREAS and ZHAireS. This contrasts with the simplified exponential refractive index model used in the ZHAireS version available when \cite{Gottowik:2017wio} was published.  Our comparisons are based on modern hadronic interaction models tuned to LHC data, which are essential for extrapolating hadronic processes to EeV energies, and are implemented in the new version of ZHAireS used here as well as in CoREAS. Also, a newer version of CoREAS is used (v1.4).

Furthermore, we extend the comparison beyond the energy emitted in radio waves, also studying the impact of using CoREAS or ZHAireS on the reconstruction of the depth of maximum (Section \ref{S:Results}), following a reconstruction methodology similar to that in \cite{Buitink:2014eqa,CARVALHO201941}.

\section{Monte Carlo Simulations of radio emission in EAS}
\label{S:Simulations}

To calculate the radiation induced by the charged particles travelling along the atmosphere in an EAS, the problem is split using the superposition principle of electromagnetism, obtaining the contributions to the electric field of the particles dividing their trajectories in numerous small rectilinear sub-tracks assuming that particles travel at a constant speed along them. Provided these sub-tracks are sufficiently small, several approximations can be made to calculate the emission \cite{Zas:1991jv,James:2010vm,Garcia-Fernandez:2012urf}. The electric field is calculated summing the contributions from all the sub-tracks of each particle trajectory, taking into account their relative phases in the frequency domain and arrival times at the observer in the time domain. This is the approach adopted in the two main programs for simulation of radio emission in EAS, CoREAS and ZHAireS, described briefly in the following.

\subsection{CoREAS}

CoREAS~\cite{Huege:2013vt} is a state-of-the-art Monte Carlo program designed to simulate radio emission from extensive air showers (EAS). It builds on the widely used CORSIKA~\cite{CORSIKA} code, which models the development of air showers in the atmosphere. CORSIKA provides detailed information for each tracked particle, including its type, energy, velocity, position, and time. CoREAS then uses this information to compute the resulting electric field using the \textit{endpoints algorithm}~\cite{James:2013nlo}, which is derived from first principles. In this algorithm, the motion of a particle along its trajectory is represented by a pair of discrete, instantaneous acceleration and deceleration events each producing radiation.

In CoREAS, the effects on radio emission resulting from the atmospheric refractive index being $n>1$ and varying with altitude with particles travelling at speeds larger than $c/n$, are taken into account in the simulation. Different models for the atmospheric refractive index can be chosen as explained in Section \ref{S:Parameters}. CORSIKA 7.7410 and CoREAS v1.4 have been used for the simulations in this work. Their respective user manuals can be consulted for further details \cite{CORSIKA_Manual, CoREAS_Manual}.

\subsection{ZHAireS}
The other state-of-the-art program for the calculation of radio emission in EAS is ZHAireS~\cite{Alvarez_Muniz_2011ref}. In this code, the AIRES shower simulation program \cite{AIRES} is used to propagate each particle in small steps that are regarded as single particle tracks. Their contribution to the radio emission is calculated and added to the total electric field in both the time~\cite{Alvarez-Muniz:2010wjm} and frequency domains~\cite{Zas:1991jv} applying the {\it ZHS algorithm} \cite{Zas:1991jv, Garcia-Fernandez:2012urf} 
originally implemented in the ZHS simulation program~\cite{Zas:1991jv, Alvarez-Muniz:2010wjm}.
This algorithm is also obtained from first principles solving Maxwell's equations. In the ZHS expression for the electric field, there are two terms for each sub-track corresponding to the start and the end points.  
In this respect, the methodology is similar to the endpoints one. A practical difference is that in the ZHS algorithm the attenuation of the electric field with distance $R$ from the source to the observer that is applied to the two terms in the expression for a sub-track, is the same \cite{Garcia-Fernandez:2012urf}, while in the endpoints formulation a different value of $R$ is independently calculated for each acceleration endpoint \cite{James:2013nlo, Alvarez-Muniz:2022uey}. In the limit of large $R$, the difference between the two approaches vanishes, but for sub-tracks that are not short compared to $R$, there can be numerical differences. 
More importantly, in the limit of observation angle with respect to the track close to the Cherenkov angle, $\theta_C$, only the ZHS approach yields a correct and finite limit~\cite{Garcia-Fernandez:2012urf}. As a result, the shower simulation programs that implement the endpoints algorithm such as CoREAS, typically adopt the ZHS expressions for sub-tracks observed at angles very close to the Cherenkov angle.

In ZHAireS, the variation with altitude of the atmospheric refractive index is modeled by default with an exponential function. Simulations in this work were performed using AIRES 19.04.08 and ZHAireS 1.1.0a, a new version\footnote{ZHAireS 1.1.0a public release is in preparation and will supersede the ZHAireS 1.0.30a version.} that incorporates additional refractive index models similar to those employed in CoREAS, as explained in Section \ref{S:Parameters}. Further details can be found in the respective AIRES and ZHAireS user manuals~\cite{AIRES_Manual, ZHAIReS_Manual}. 

\subsection{Parameters and settings of the CoREAS and ZHAireS simulations}
\label{S:Parameters}
%

To perform the comparisons presented in this article, sets of shower simulations with the same primary particle, energy, geometry, magnetic field, hadronic interaction model and refractive index model were performed with both CoREAS and ZHAireS. 

\begin{figure}
    \centering
    \includegraphics[width=\linewidth]{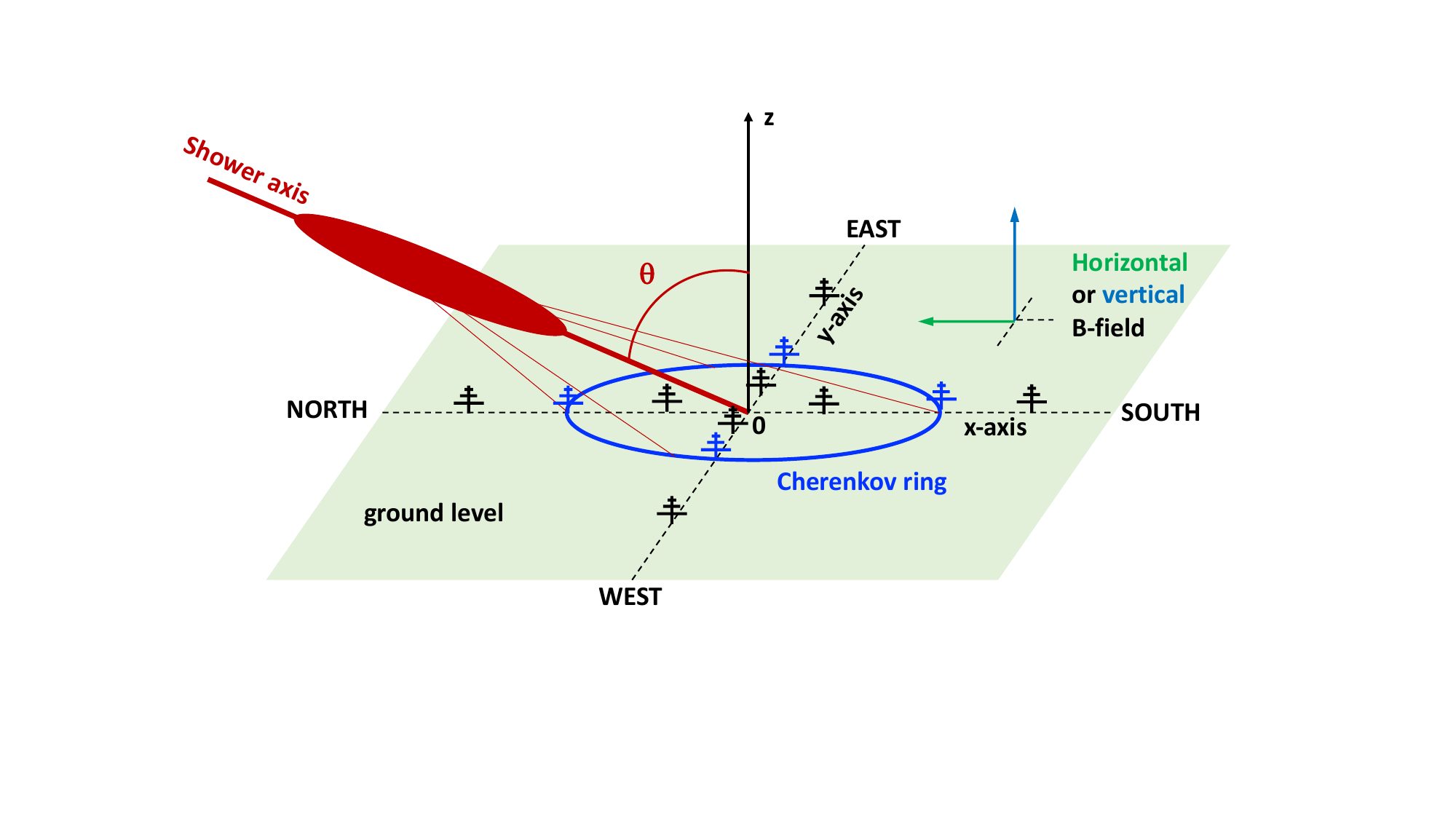}
    \vspace{-2.5cm}
    \caption{Sketch of the coordinate system used in this work defining north, south, east, and west in relation to the Earth's magnetic field orientation. The configurations of the magnetic field (horizontal or vertical) adopted in this study are displayed (green and light blue arrows). The antenna positions typically along the north-south ($x$--axis) and east-west ($y$--axis) direction, unless otherwise indicated, are also shown. The shower axis, by default along the north-south direction for showers arriving at the ground from north, and the zenith angle $\theta$ are indicated. The curvature of the Earth, not shown in the sketch, is accounted for in the simulations. The \textit{Cherenkov ring} of the radio signal, defined in the text, is represented as a blue ellipse.}
    \label{fig:esquema}
\end{figure}

Fig.\,\ref{fig:esquema} depicts the coordinate system and geometrical parameters used in this work. The results of the simulations presented in the following correspond to particle showers that propagate along the north-south direction (moving southward and originating from the north), as shown in Fig.\,\ref{fig:esquema}. Observers (antennas) are typically placed along the north-south ($x$--axis) and east-west ($y$--axis) directions unless otherwise indicated. Simulations are performed with either a horizontal magnetic field, represented by a green line (parallel to the ground and pointing north), or a vertical magnetic field, represented by a blue line (perpendicular to the ground). Although the geomagnetic field is generally neither purely horizontal nor vertical, it can always be decomposed into horizontal and vertical components. These idealized configurations are therefore physically meaningful and offer intuitive, easy-to-visualize $(\vec{v} \times \vec{B})$ and $(\vec{v} \times (\vec{v} \times \vec{B}))$ directions on the ground, thereby simplifying the analysis.

Other simulation parameters and settings, including energy thresholds below which particles are no longer tracked, were carefully chosen to reduce potential intrinsic discrepancies arising from shower development. We briefly discuss below several of these parameters and settings for reproducibility of our results. 

\begin{itemize}

\item 
Atmospheric refractive index model:
ZHAireS and CoREAS take into account the fact that the atmospheric refractive index $n$ is larger than one and decreases with altitude $h$ above ground.
CoREAS can model the refractive index with the Gladstone-Dale (GD) law which assumes that $n(h)$ varies in proportion to the atmospheric density. This model uses as input the value of the refractive index at ground and the density profile of the atmosphere. By default, ZHAireS uses an exponential model for $n(h)$ that is parameterized in terms of the refractivity ${\cal R}(h)=[n(h)-1]\times 10^6$ and given by ${\cal R}(h)={\cal R}_s \exp{(-{\cal K}_r h)}$, where ${\cal R}_s$ and ${\cal K}_r$ can be changed by the user.
However, since version 1.1.0a the Gladstone-Dale law is also implemented in ZHAireS and used in this work. 
To ensure consistency, the same atmospheric density profile (US Standard Atmosphere) was used in both CoREAS and ZHAireS.

\item

The \texttt{STEPFC} parameter:
CORSIKA uses the EGS4 simulation package \cite{Nelson:1985ec} for the transport of electrons, positrons and photons. A parameter dubbed the \texttt{STEPFC} factor allows to change the step length in the multiple scattering of electrons and positrons.  By reducing \texttt{STEPFC} a finer simulation of the particle cascade is achieved \cite{Gottowik:2017wio}, affecting observables such as the energy released in the form of radio waves that can increase by $\sim 10\%$ when reducing the value of \texttt{STEPFC}. The default value in CORSIKA is \texttt{STEPFC\,=\,1}. ZHAireS uses a different tracking algorithm that does not include a tunable parameter like \texttt{STEPFC}. However, it was estimated in~\cite{Gottowik:2017wio} that the default ZHAireS simulations correspond to an effective step parameter of \texttt{STEPFC} $\simeq$ \texttt{0.07}. To achieve the best possible accuracy in the comparisons of the two simulations, and as recommended in \cite{Gottowik:2017wio}, \texttt{STEPFC} $=$ \texttt{0.05} was used in the CORSIKA-CoREAS simulations in this study. 

\begin{figure}[ht]
    \centering
      {\includegraphics[width=0.48\linewidth]{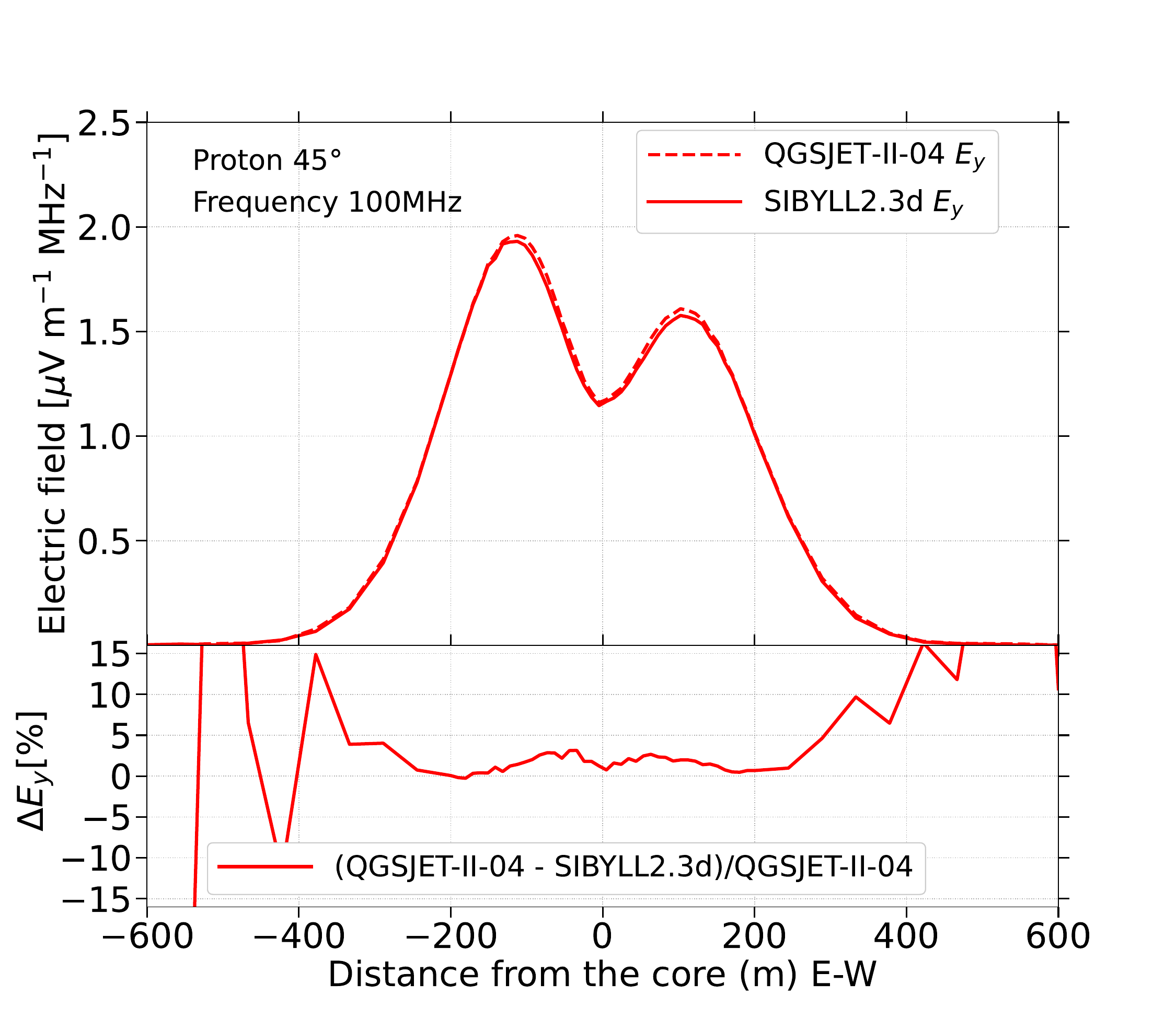}}
      {\includegraphics[width=0.48\linewidth]{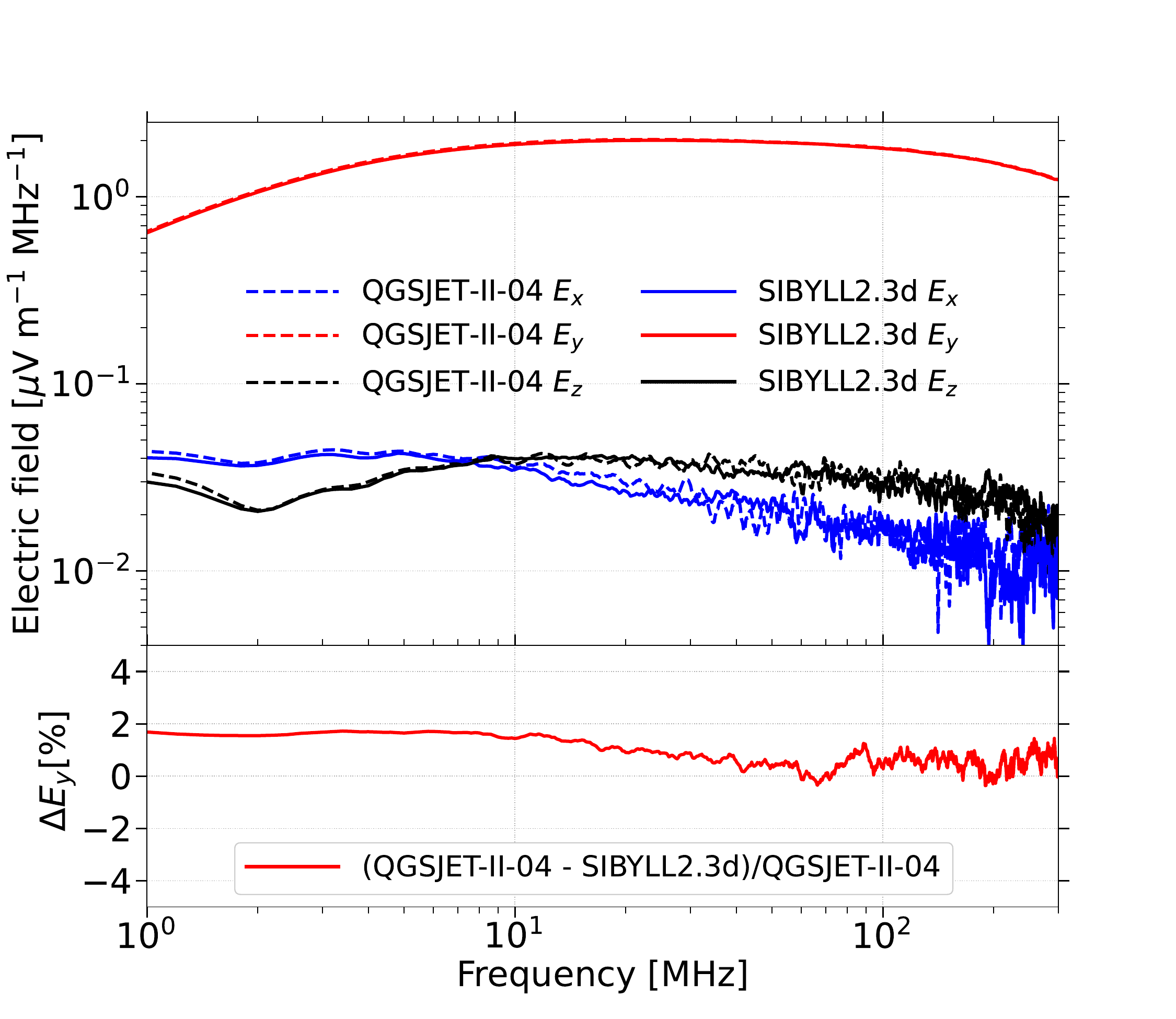}}
   \caption{Electric field obtained in ZHAireS simulations of primary protons with energy \(E = 10^{17} \, \text{eV}\) and zenith angle \(\theta = 45^\circ\), adopting SIBYLL 2.3d (solid lines) and QGSJETII.04 (dashed lines) as hadronic models.
    The showers were selected so that their \(X_{\rm max}\) values are similar: \(X_{\rm max}(\text{SIBYLL}) = 684.2 \, \text{g/cm}^2\) and \(X_{\rm max}(\text{QGSJET}) = 684.1 \, \text{g/cm}^2\). A horizontal magnetic field configuration was used, with \( |\vec{B}| = 50 \, \mu\text{T} \), parallel to the ground and pointing towards magnetic north. 
   Left panel: Modulus of the $E_y$ component of the Fourier transform of the electric field at frequency 100 MHz for observers located at different distances from the shower core along the east-west (E-W) direction (see Fig.\,\ref{fig:esquema}). 
   Positive coordinates correspond to observes east of the shower core. The showers develop from north to south.  
   Right panel: the frequency spectrum of the Fourier components of the electric field $E_{x}$ (blue), $E_{y}$ (red), and $E_{z}$ (black) for observers on the ground at a distance $r\simeq 150\,$m west from the shower core near the {\it Cherenkov ring} (see Fig.\,\ref{fig:esquema}). 
   In the bottom panel of each plot we show the relative difference (QGSJET$-$SIBYLL)/QGSJET of the $E_y$ component. } 
   \label{fig:ZHAireS-had-models}
\end{figure}

\item 
Hadronic interaction models:
For CORSIKA, we adopted the high-energy hadronic interaction model QGSJETII.04~\cite{Ostapchenko:2010vb} tuned to LHC data. The more recent version of ZHAireS used in this work, not available at the time the study in \cite{Gottowik:2017wio} was performed, also incorporates more modern versions of hadronic interaction models tuned to LHC data, and in particular the model QGSJETII.04. Other possibilities include SIBYLL2.3d\cite{Riehn:2019jet} and EPOS-LHC~\cite{Pierog:2013ria}. 

We have studied the influence of different hadronic models on radio emission by simulating a set of EAS using only ZHAireS, with the hadronic models QGSJETII.04 and SIBYLL2.3d. In the left panel of Fig.~\ref{fig:ZHAireS-had-models}, the dominant Fourier component of the electric field $E_{y}$ (parallel to $\hat v \times \vec B$) as a function of distance to the shower core (see Fig.\,\ref{fig:esquema} for the geometry) is presented, as predicted in ZHAireS simulations, for proton primaries and zenith angle $\theta=45^\circ$ using the two chosen hadronic interaction models.  
The relative differences between the electric field $E_y$ as predicted by both models are also shown in the same figure. These are below $\simeq 2\%$ in the range of distances to the core where the electric field is at most a factor $\sim 4$ smaller than the peak value. 

In the right panel of Fig.~\ref{fig:ZHAireS-had-models}, we show the Fourier components $E_{x}$, $E_{y}$, and $E_{z}$ of the electric field  plotted against frequency as predicted by ZHAireS simulations. The observers were positioned close to the \textit{Cherenkov ring}\footnote{The location of this \textit{ring} can be estimated by considering that the dominant contribution to the radio signal arises at the altitude corresponding to the depth of the shower maximum ($X_{\rm max}$) and is emitted at the Cherenkov angle relative to the shower axis.}, where the highest emission is expected. The differences between the frequency spectrum of the dominant $E_y$ component for the different hadronic models remain below $2\%$ for frequencies between 1 MHz and 300 MHz.

We have also verified that these relative differences are not due to the differences in the fraction of energy into the electromagnetic component of the shower, as predicted by the two different hadronic models. Specifically, for the showers simulated using QGSJETII.04 approximately 11.1 \% of the primary energy is carried by muons and neutrinos on average, whereas for those simulated with SIBYLL2.3d this fraction is approximately 11.5 \%, corresponding to an approximate difference in electromagnetic energy $\lesssim 0.5\%$.

Predictions of the energy emitted in the form of radio waves using alternative hadronic interaction models, such as EPOS-LHC in comparison to QGSJetII-04, were obtained in \cite{Glaser:2016qso}, concluding that their impact is also negligible.

\item 
Particle energy thresholds:
For reproducibility of our results, we give in Table\,\ref{table:thinning} the values of the kinetic energy thresholds below which particles of different species are discarded in the simulations. We have used the same values of the energy thresholds in both ZHAireS and CoREAS simulations.

\begin{table}[ht]
\scriptsize
\begin{center}
\renewcommand{\arraystretch}{1.3}
\begin{tabular}{ l l }
\hline
{\bf AIRES / ZHAireS} & \hspace{1cm}{\bf CORSIKA / CoREAS} \\ 
\hline
{\texttt{NuclCutEnergy    300 MeV}} & \hspace{1cm}{\texttt{ECUTS   3.000E-01 1.000E-02 2.50E-04 2.5E-04}} \\ 
{\texttt{MuonCutEnergy     10 MeV}}    &                 \hspace{1cm} ~~~(GeV)~~~~~~~Hadrons~~~~~~~~~Muons~~~~~~~~~Electrons~~~~~~~Gammas \\ 
{\texttt{ElectronCutEnergy 250 keV}} & \\
{\texttt{GammaCutEnergy    250 keV}} & \\
\hline
{\texttt{ThinningEnergy 1.e-6 Relative}} & \hspace{1cm}{\texttt{THIN    1.000E-06 1.000E+02 50.0E+02}}  \\ 
{\texttt{ThinningWFactor 0.06}}     & \hspace{1cm}{\texttt{THINH   1.000E+00 1.000E+02}} \\ 
\hline
\end{tabular}
\caption{Particle kinetic energy thresholds (rows 1\,--\,4) and thinning options (5 and 6) used in the CORSIKA/CoREAS, and AIRES/ZHAireS shower simulations in this work. For AIRES/ZHAireS the particle species to which the energy threshold applies is indicated in the name of the input directive. In CORSIKA/CoREAS the energy threshold affects hadrons, muons, electrons and gammas respectively.}
\label{table:thinning}
\end{center}
\end{table}

\item 
Thinning parameters: 
The thinning algorithms used in EAS simulations to reduce computational time are implemented differently in CORSIKA/CoREAS and AIRES/ZHAireS, despite being based on similar principles. These algorithms aim to track only a small, representative fraction of the particles in a shower that fall below a certain energy threshold (usually a fraction of the primary energy). Each tracked particle is then assigned a weight to compensate for the particles that are discarded. A maximum weight is allowed in both simulations. In AIRES/ZHAireS a single thinning level is used for all particle types \cite{AIRES_Manual}, while in CORSIKA/CoREAS different thinning levels can be set for electromagnetic and hadronic particles \cite{CORSIKA_Manual,Kobal:2001jx}. 
Using the default thinning parameters for both simulations leads to significantly different settings, particularly in the maximum allowed weight. To ensure consistency, we have fixed the relative thinning and weight limitation factor to the values listed in Table~\ref{table:thinning}. This choice ensures that the maximum weight remains approximately the same in both simulations.

\item 
Curvature of the Earth:
In AIRES/ZHAireS, the curvature of the Earth is taken into account by default, and showers with zenith angles up to and even above $\theta=90^\circ$ can be reliably simulated \cite{Tueros:2023ifq}.
Similarly, in CORSIKA/CoREAS, the curvature of the Earth can be taken into account activating the option \texttt{CURVED}, which we have adopted in this work for the simulation of showers with $\theta>70^\circ$. 
Since a full spherical calculation is computationally expensive, different strategies and approximations are adopted in both CORSIKA and AIRES. See the respective manuals \cite{CORSIKA_Manual, AIRES_Manual} for further details.

\end{itemize}

\subsection{Reduction of shower-to-shower fluctuations}
\label{S:Methodology}


To enable accurate comparisons between the properties of radio emission in CoREAS and ZHAireS, we tried to minimize the impact of intrinsic differences stemming from the shower development and their inherent shower-to-shower fluctuations. To achieve this, the following specific procedure was adopted:

\begin{itemize}

\item 
First, a large set of air showers with identical primary particle, primary energy, zenith and azimuth angles, was simulated using CORSIKA and AIRES with the relevant settings explained in Section\,\ref{S:Parameters}, excluding the calculation of radio emission in this first stage. From these simulations, 20 showers from each set were selected, ensuring their depths of shower maximum ($X_\text{max}$) were all within $1\,\mathrm{g/cm^2}$ of a specified value.

\item 
Once these showers were selected, simulations were rerun with CoREAS and ZHAireS, now including the calculation of radio emission, using the same geomagnetic field configuration and other parameters relevant to radio emission in both cases. Importantly, enabling the radio emission calculation in ZHAireS and CoREAS did not alter the sequence of random numbers in the simulation, ensuring the originally selected $X_\text{max}$ values remained unchanged.

\end{itemize}

This approach eliminates the need to correct the geomagnetic and Askaryan contributions to the radio emission for the geomagnetic angle and atmospheric density at $X_\text{max}$, allowing for a more direct comparison between the two simulation codes.

\section{Results}
\label{S:Results}

Using the methodology outlined in Section\,\ref{S:Methodology}, we have compared the radio emission of air showers simulated with CoREAS and ZHAireS. Our analysis focuses on the electric field components in both the frequency and time domains, the energy emitted as radio waves and its spatial distribution in the shower plane, as well as on the influence of the simulation code on the reconstructed depth of maximum shower development based on the information provided by the radio signal.

\subsection{Electric field}

We adopt the coordinate system shown in Fig.\,\ref{fig:esquema}, with the magnetic field either parallel to ground or perpendicular to it, and the showers always arriving from magnetic North. In this manner, the $y$-component $E_y$ of the electric field is parallel to $\vec{v}\times\vec{B}$ and hence dominated by geomagnetic emission. 

\begin{figure}[ht]
  \centering
      {\includegraphics[width=0.48\linewidth]{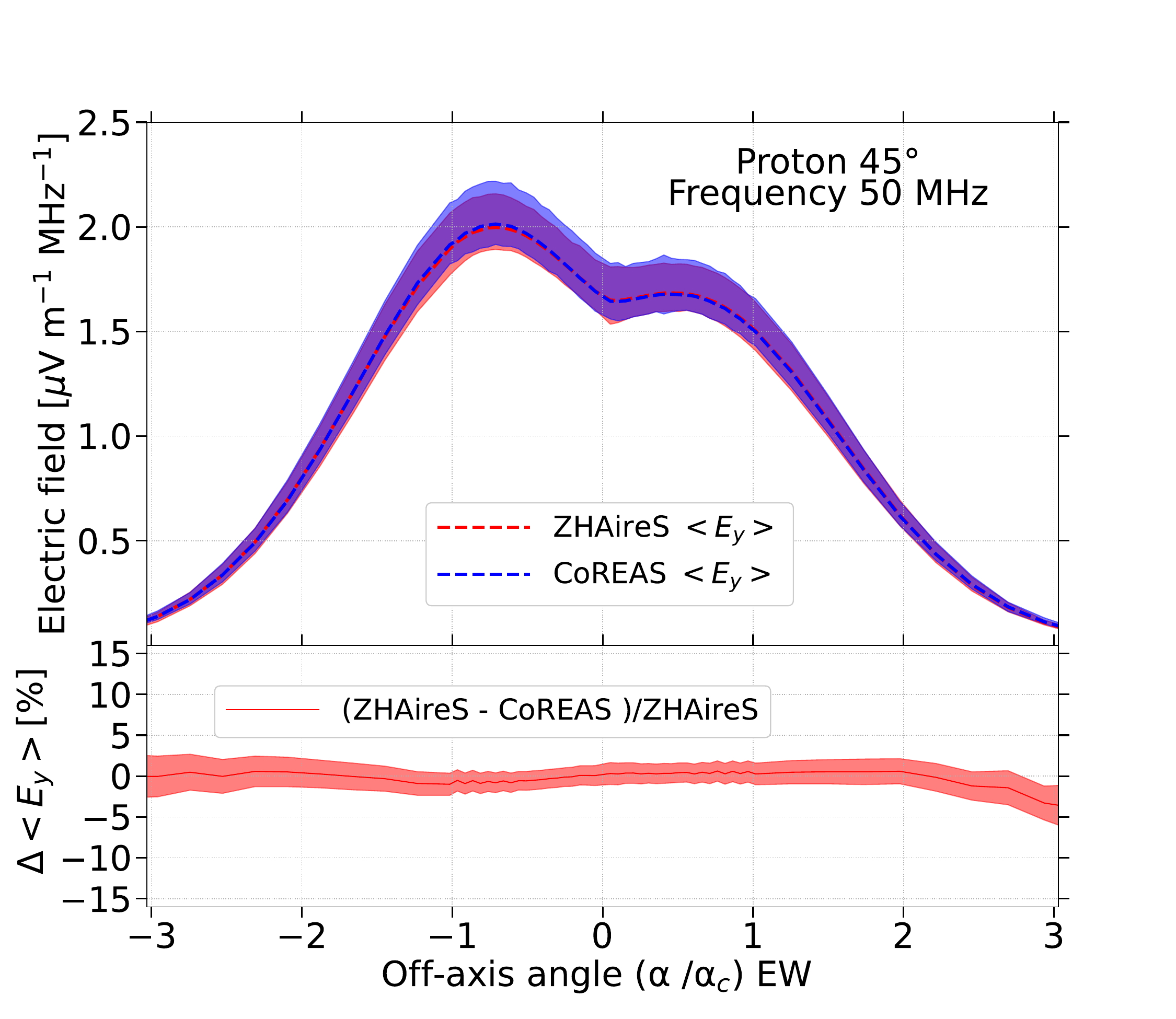}}
      {\includegraphics[width=0.48\linewidth]{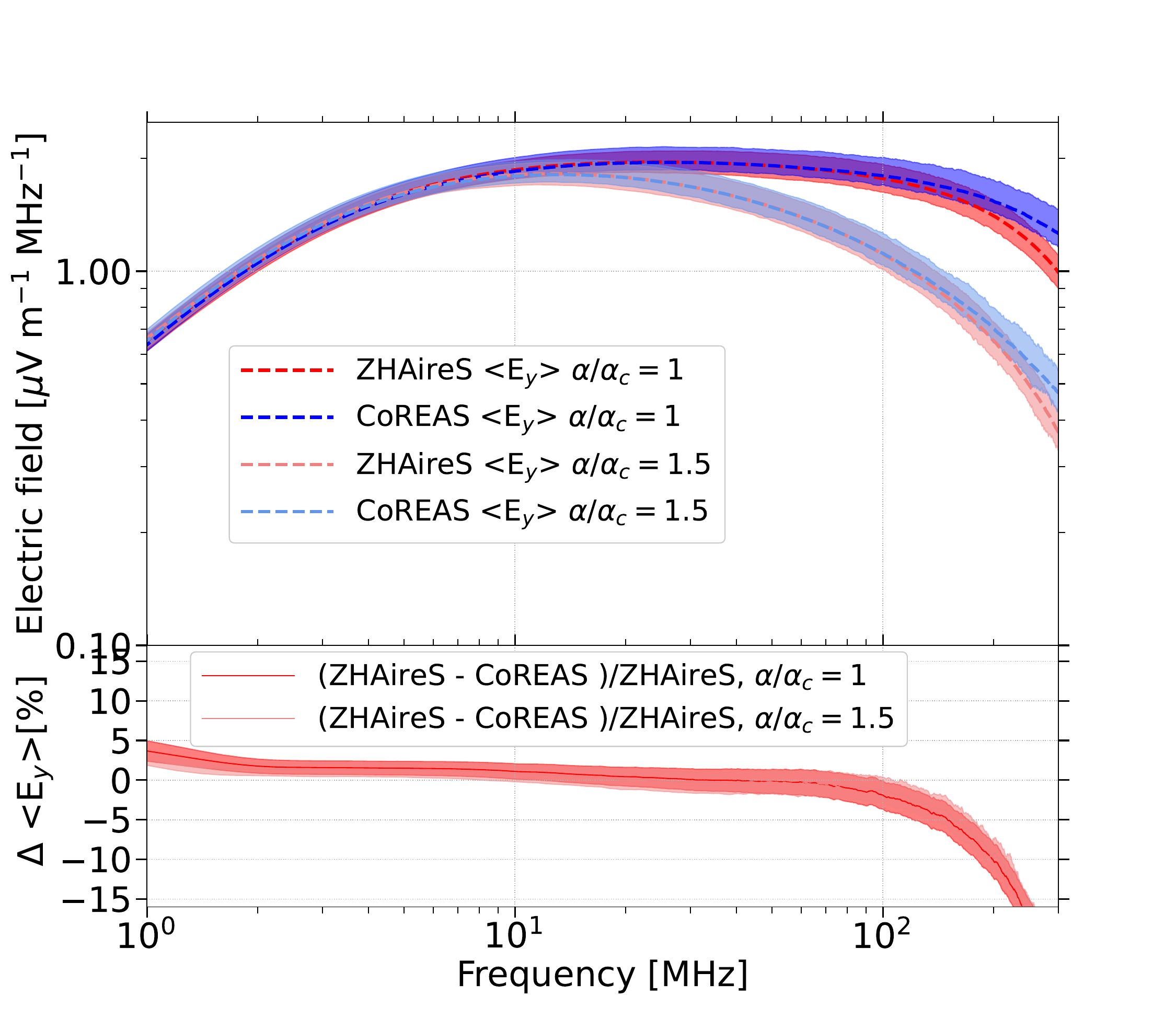}}
   \caption{Modulus of the Fourier component $E_y$ of the electric field as obtained in simulations performed with ZHAireS and CoREAS, 20 showers each, for primary protons with $E=10^{17}$ eV and $\theta=45^\circ$. The showers were selected so that their $X_{\rm max}$ values are all within less than $1\,\mathrm{g\,/cm^2}$ of $X_{\rm max}=642.5 \, \mathrm{g/cm^2}$. We used a horizontal magnetic field configuration with $|\vec{B}|=50\,\mu$T, parallel to the ground and pointing towards magnetic north. The dashed lines correspond to the average $\langle E_{y}\rangle$ while the shaded area represents the region between $\mathrm{max}(E_{y})$ and $\mathrm{min}(E_{y})$ for ZHAireS (red) and CoREAS (blue) over the 20 simulations. 
   Left panel: $E_y$ component at 50 MHz as a function of the off-axis angle $\alpha$ with respect to the Cherenkov angle $\alpha_C$. The observers are located along the east-west (EW) direction (see Fig.\,\ref{fig:esquema}). Positive coordinates correspond to antenna positions located east of the core. The showers develop from north to south. 
   Right panel: $E_y$ for observers located at distances to the east of the shower core on the ground, corresponding to off-axis angles $\alpha/\alpha_C\simeq 1$ and $\alpha/\alpha_C\simeq 1.5$.
   The bottom panel of each plot shows the relative difference between the averages, $(\rm ZHAireS - \rm CoREAS)/\rm ZHAireS$, (solid line). The shaded band represents the $1\,\sigma$ spread propagated to the relative difference.}
   \label{fig:proton45LDF1}
\end{figure}

\begin{figure}[ht]
  \centering
      {\includegraphics[width=0.48\linewidth]{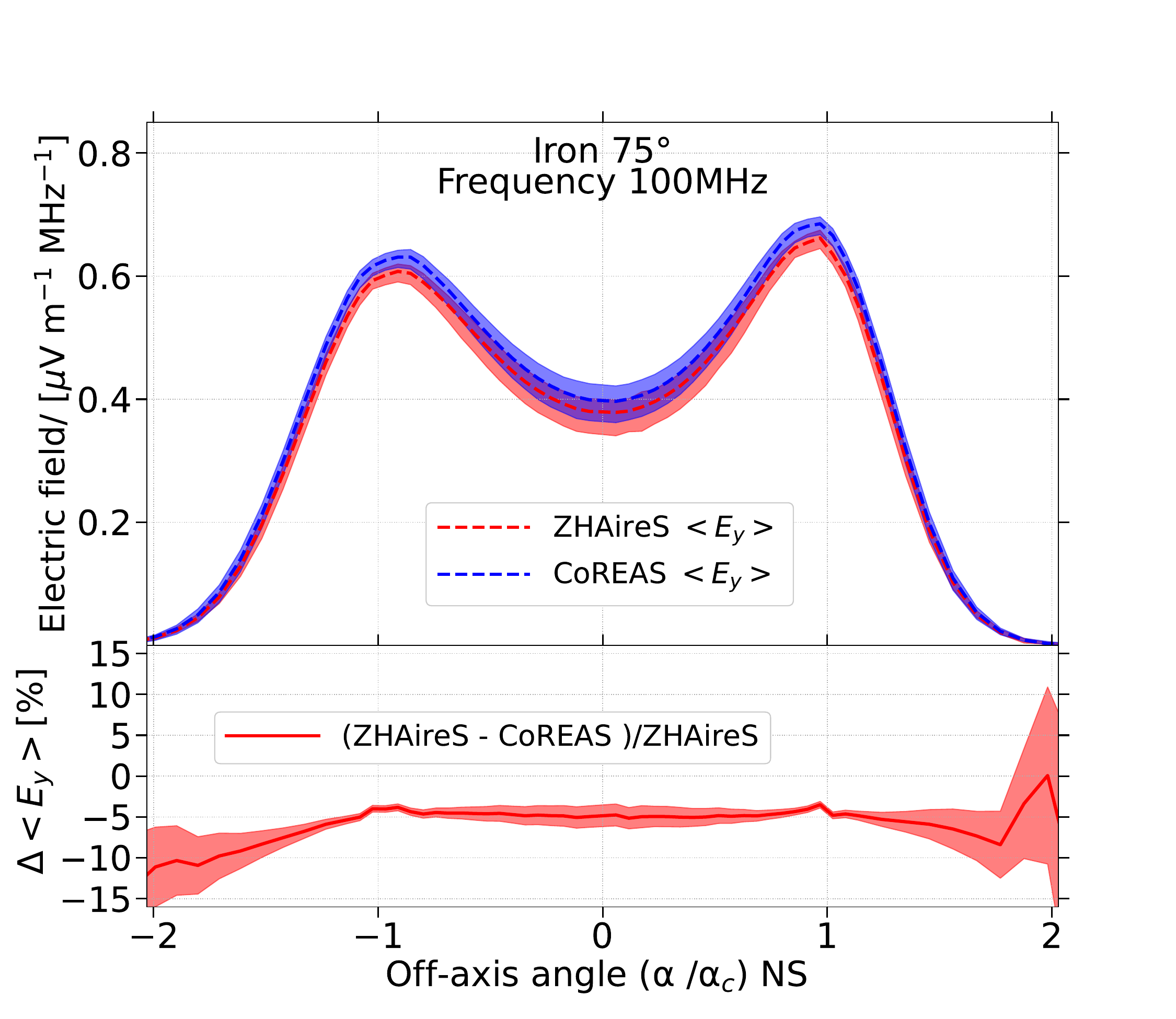}}
      {\includegraphics[width=0.48\linewidth]{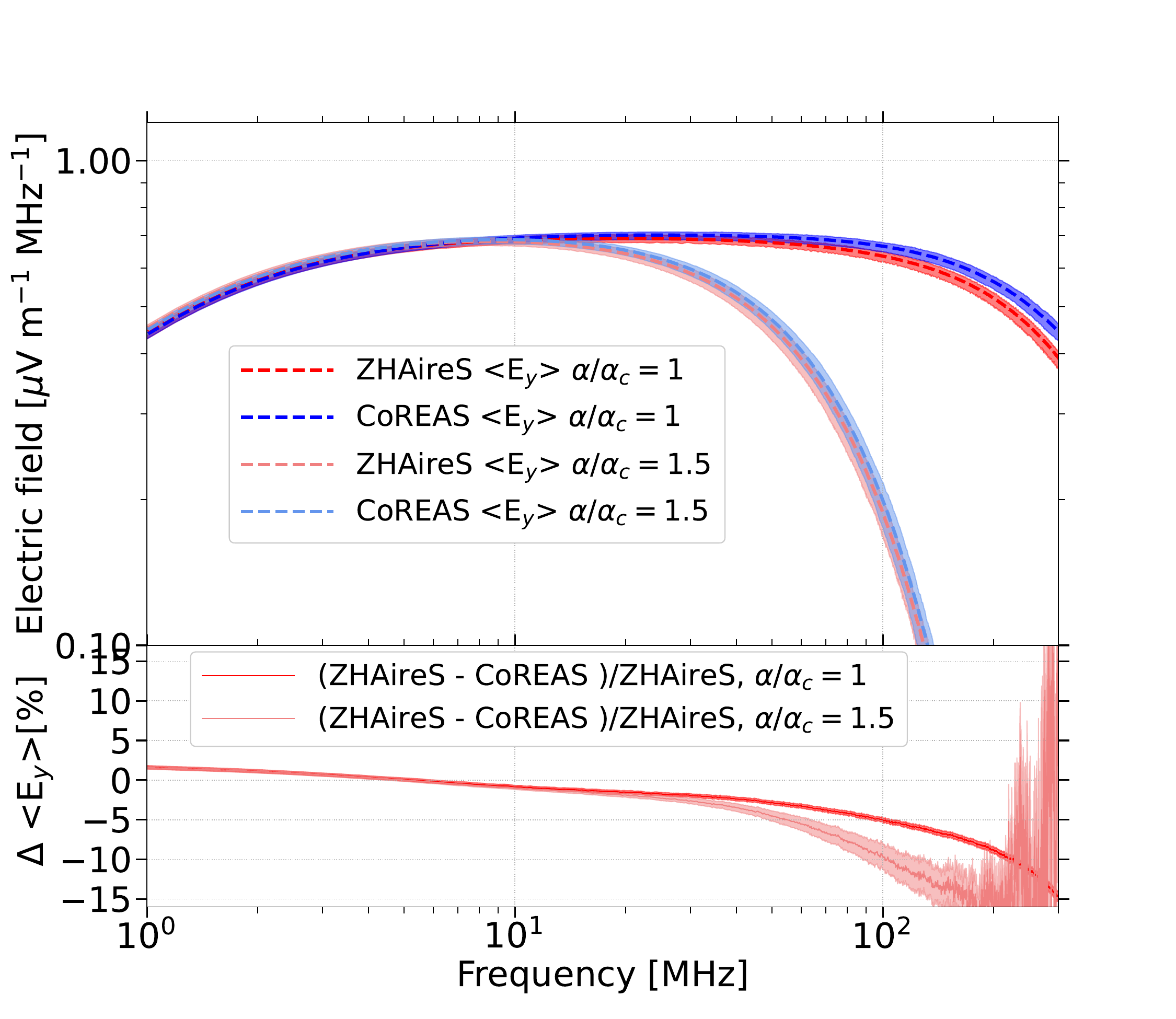}}
  \caption{Same as Fig.\,\ref{fig:proton45LDF1} 
for primary iron nuclei with $E=10^{17}$ eV and $\theta=75^\circ$ and a vertical magnetic field configuration with $|\vec{B}|=50\,\mu$T, perpendicular to the ground. 
The simulated showers were selected so that their values of $X_{\rm max}$ are within less than $1\,\mathrm{g/cm^2}$ of $X_{\rm max}=574.5 \, \mathrm{g/cm^2}$. In the left panel the observers are located along the north-south (NS) direction (see Fig.\,\ref{fig:esquema}), with north corresponding to the positive x-axis. In the right panel the antennas are placed at positions north of the shower core. 
} 
   \label{fig:iron75LDF1}
\end{figure}

In Fig.~\ref{fig:proton45LDF1}, 
the Fourier components of the $E_y$ polarization of the radio emission predicted in CoREAS and ZHAireS are compared for showers with zenith angle $\theta=45^\circ$ arriving from magnetic north, induced by primary protons with energy $E=10^{17}$ eV. A horizontal magnetic field with total intensity $|\vec{B}|=50\,\mu$T  is used in the simulations (see Fig.\, \ref{fig:esquema}), corresponding to geomagnetic angle, $\beta$, between the shower axis and $\vec B$ with $\sin\beta=0.7$.
For this comparison, 20 proton-induced showers were simulated having similar $X_\text{max}$ values within $< 1\,\mathrm{g/cm^2}$ of the average value of the sample $\langle X_\mathrm{max}\rangle=642.4 \, \mathrm{g/cm^2}$. Similarly, in Fig.~\ref{fig:iron75LDF1} 
the predicted emission is compared for inclined showers induced by primary iron with an energy of $E=10^{17}$ eV, $\theta=75^\circ$ and a vertical magnetic field of intensity $|\vec{B}|=50\,\mu$T (see Fig.\,\ref{fig:esquema}), so that for this specific case $\sin\beta=0.97$. 
Again 20 showers were simulated, in this case with $X_\mathrm{max}$ values within $\simeq 1\,\mathrm{g/cm^2}$ of the average $\langle X_\mathrm{max}\rangle = 574\, \mathrm{g/cm^2}$.

The dashed lines in the figures represent the average value $\langle E_y \rangle$ of the 20 simulations, while the filled areas indicate the region between the maximum ($\max(E_y)$) and minimum ($\min(E_y)$) values over the 20 simulations for ZHAireS (red) and CoREAS (blue). The bottom panels illustrate the relative difference between the average values of $E_y$ for ZHAireS and CoREAS, calculated as $(\text{ZHAireS}-\text{CoREAS})/\text{ZHAireS}$.

Firstly, and as expected, for the same primary energy the amplitude of the $E_y$ component in the more vertical $\theta=45^\circ$ proton-induced showers is approximately three times larger than that in the $\theta=75^\circ$ iron-induced showers. This is mainly due to the increased geometrical distance from ground to $X_{\rm max}$ due to the larger inclination of the shower and the smaller value of $X_{\rm max}$ of the iron showers, 
with the compensating effect of the different geomagnetic angles ($\sin\beta=0.7$ for $\theta=45^\circ$ vs $\sin\beta=0.97$ for $\theta=75^\circ$) playing a secondary role. 

\subsubsection{Lateral distribution}

In the left panel of Fig.~\ref{fig:proton45LDF1} (Fig.\,\ref{fig:iron75LDF1}), the dominant $E_y$ component is shown at a frequency of 50 MHz (100 MHz) and as a function of the distance from the shower axis, expressed in terms of the off-axis angle corresponding to the angular distance from the shower axis in units of the Cherenkov angle. 
The observers in Fig.\,\ref{fig:proton45LDF1} are located along the east-west (EW) direction, while those in Fig.\,\ref{fig:iron75LDF1} are placed along the north-south (NS) direction. 

The lateral distribution of the radio signal predicted by CoREAS and ZHAireS exhibits a similar shape, with a distinct \textit{Cherenkov ring}, especially at higher frequencies. This ring corresponds to the elliptical region on the ground where the emission reaches its maximum. The ring is clearly observed through two peaks in the Fourier component $E_y$ of the electric field. As is well known, these peaks are not symmetric along the east-west direction because, in the east, the geomagnetic component roughly adds to the Askaryan effect, while in the west, they subtract \cite{PierreAuger:2014ldh,Belletoile:2015rea,Schellart:2014oaa}. The asymmetry is expected to be much smaller in the north-south direction, where the Askaryan and geomagnetic polarizations are orthogonal, as illustrated in Fig.\,\ref{fig:iron75LDF1}. 

Inspecting Figs.\,\ref{fig:proton45LDF1} 
and \ref{fig:iron75LDF1}, the differences between the predictions of the two simulation programs are $\lesssim 5\%$ when normalized to the ZHAireS value and for radial distances to the core corresponding to off-axis angles $-1.5\lesssim\alpha/\alpha_C\lesssim1.5$. These differences typically increase for larger distances, and are slightly larger at 100 MHz than at 50 MHz. In all cases, ZHAireS predicts systematically smaller values of the electric field than CoREAS. These findings hold for various distances from the shower core along the EW and NS directions and for different frequencies. 

\begin{figure}[ht]
  \centering
  \begin{minipage}{0.9\linewidth}
    \centering
      {\includegraphics[width=\linewidth]{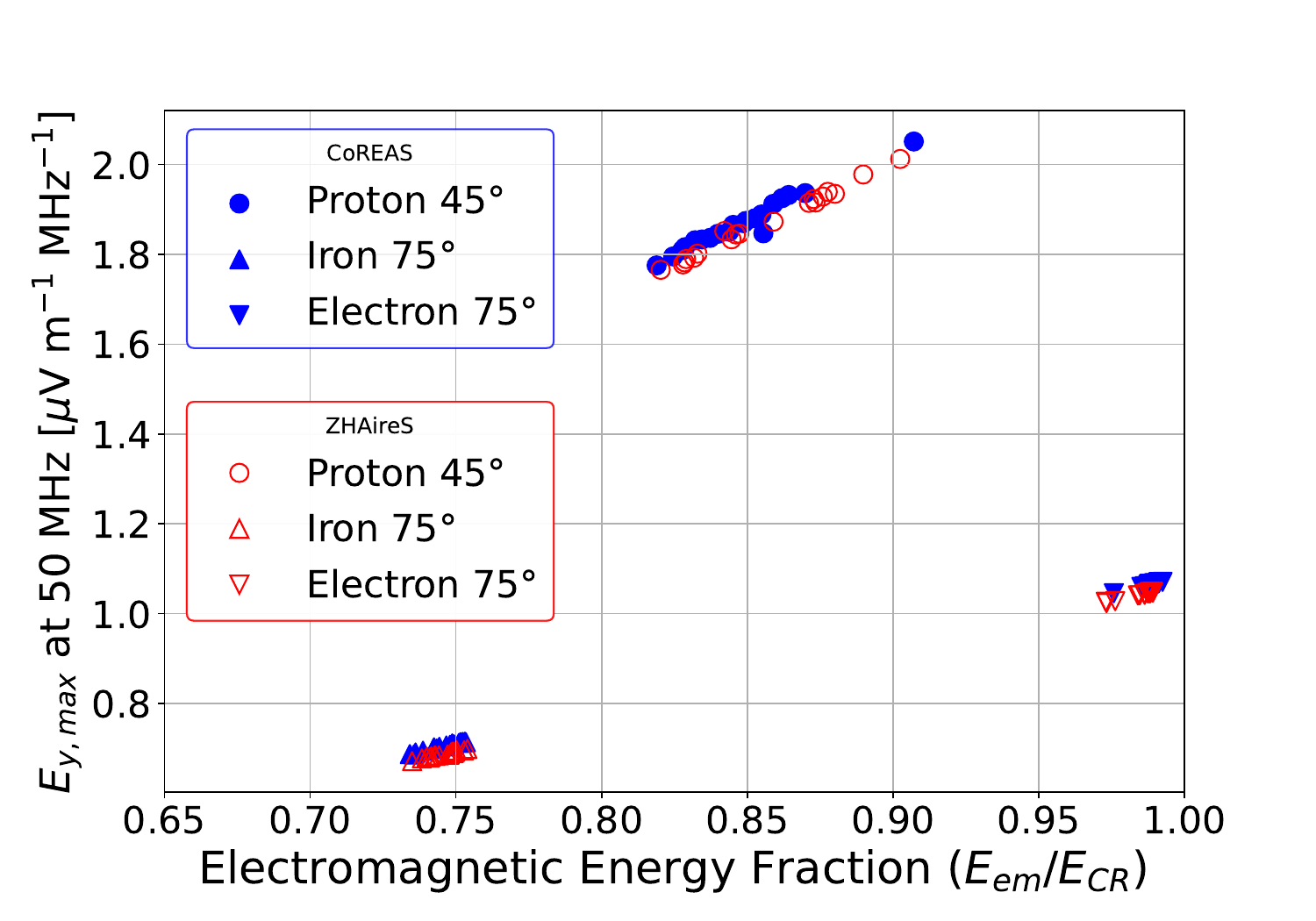}}
  \end{minipage}
\caption{Peak value of the $y$-component of the electric field ($E_{y,\mathrm{max}}$) at a frequency of 50 MHz as a function of the fraction of the primary energy ($E=10^{17}$ eV) in electromagnetic particles  (see text for definition) for protons with $\theta=45^\circ$ (circles), iron $\theta=75^\circ$ (triangles), and electrons $\theta=75^\circ$ (inverted triangles). Filled blue symbols: CoREAS simulations; empty red symbols: ZHAireS simulations.
}
  \label{fig:correlation}
\end{figure}

The spread of the signal around the average value, quantified as $\max(E_y) - \min(E_y)$, is also similar in the CoREAS and ZHAireS simulations. This variation is primarily attributed to the different fractions of electromagnetic energy carried by each shower in the sample, as the electric field is predominantly generated by electrons and positrons. 
To test this hypothesis, we present in Fig.\,\ref{fig:correlation} the maximum of the electric field component $E_y$ at 50 MHz, plotted as a function of the fraction of electromagnetic energy of the shower\footnote{We define the electromagnetic energy as the sum of three terms: the energy deposited by $e^-+e^+$ in the atmosphere; the energy of $e^-+e^+$ and photons below threshold discarded in the simulations; and the energy of $e^-+e^++\gamma$ reaching ground.}.  
As expected, an almost linear correlation is observed in both codes and for proton and iron primaries, 
with showers containing a larger fraction of electromagnetic energy emitting stronger electric fields. 
The spread is significantly larger in proton than iron simulations, consistent with the expected smaller shower-to-shower fluctuations of the electromagnetic energy in iron-induced showers\,\cite{Barbosa:2003dc}. 

\begin{figure}[ht]
  \centering
      {\includegraphics[width=0.48\linewidth]{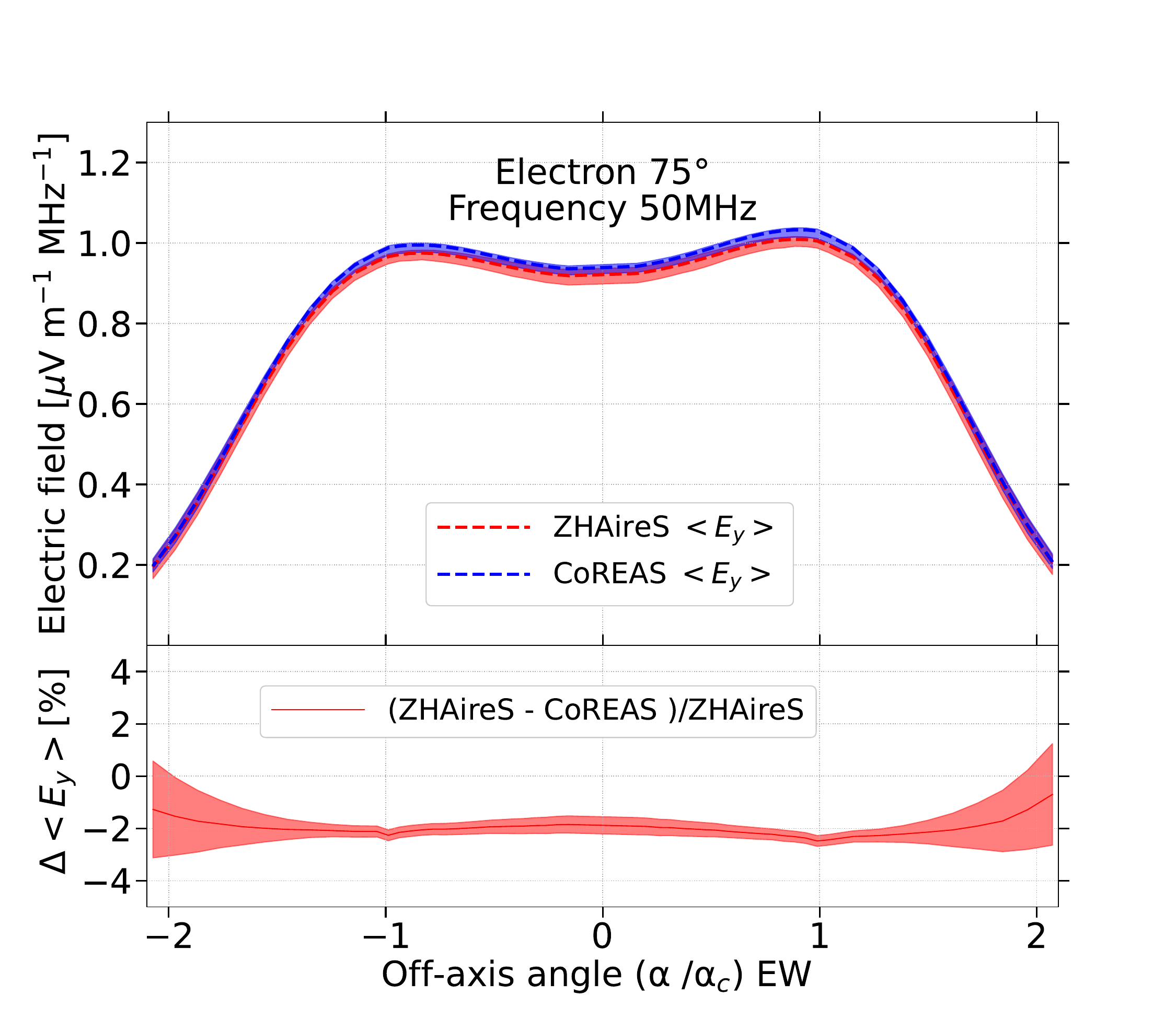}}
      {\includegraphics[width=0.48\linewidth]{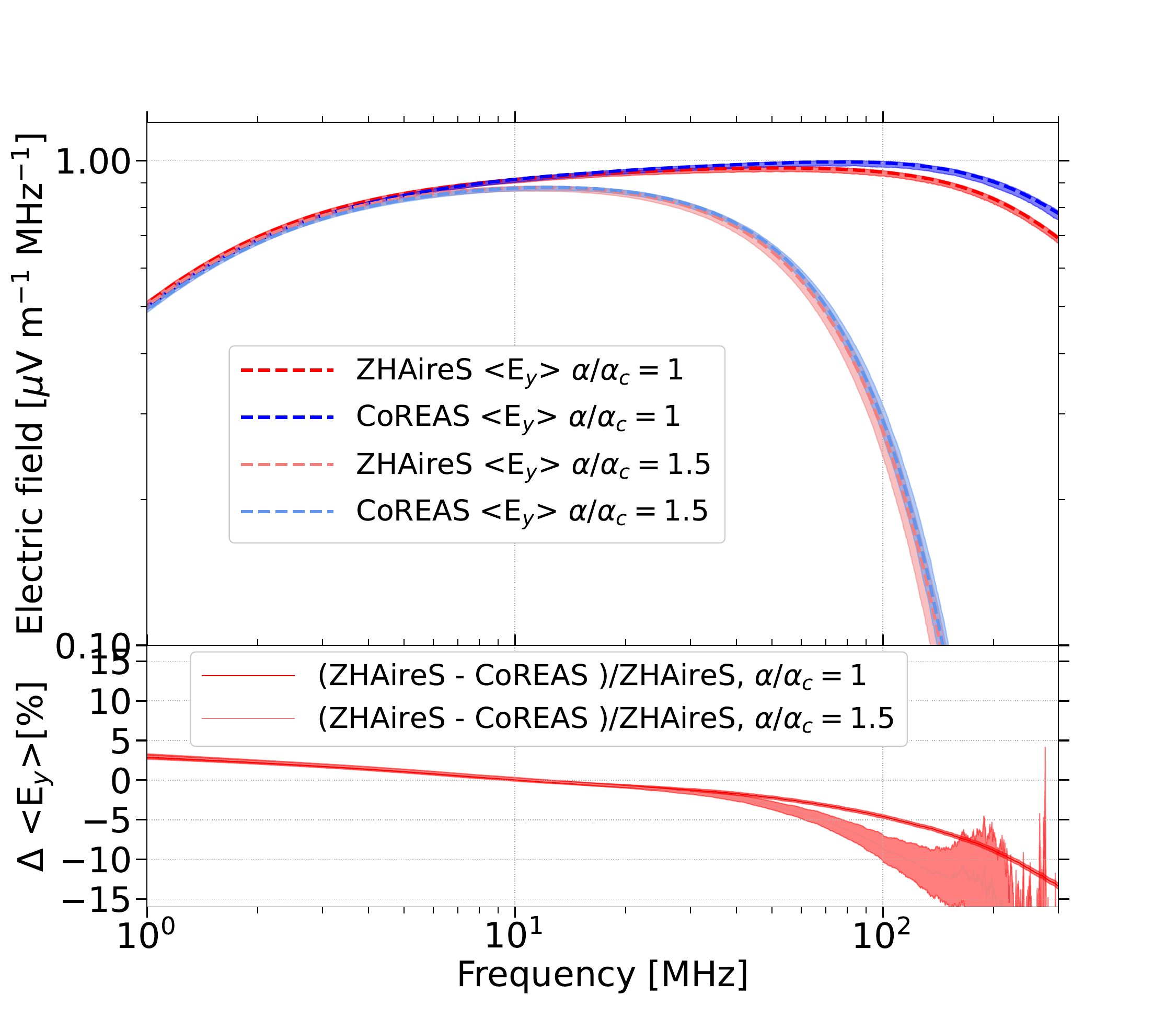}}
\caption{Same as Fig.\,\ref{fig:proton45LDF1} 
for primary electrons with $E=10^{17}$ eV and $\theta=75^\circ$ and a vertical field configuration with $|\vec{B}|=50\,\mu$T, perpendicular to the ground. The 20 simulated showers with each CoREAS and ZHAireS were selected so that their values of $X_{\rm max}$ are within less than $1\,\mathrm{g/cm^2}$ of $X_{\rm max}=715.2 \, \mathrm{g/cm^2}$. 
} 
   \label{fig:electron75LDF1}
\end{figure}

To further validate this interpretation, 20 electron-induced showers of energy $E=10^{17}$ eV and $\theta=75^\circ$ were simulated. These showers are expected to exhibit a smaller spread in electromagnetic energy compared to proton or iron-induced showers, with a significantly lower fraction of non-electromagnetic particles. The resulting electric fields at 50 MHz are also shown in Fig.\,\ref{fig:correlation} where the smaller spread and larger electromagnetic energy content is apparent. It can also be seen that regardless of the electromagnetic energy fraction, ZHAireS consistently gives a slightly lower electric field amplitude than CoREAS ranging from $2\%$ to $5\%$ (using ZHAireS as normalization) depending on the primary particle. Interestingly, the fraction of shower energy carried by the electromagnetic component is nearly identical in CoREAS and ZHAireS, indicating that the observed differences in electric field normalization do not stem from discrepancies in the electromagnetic energy predicted by the two codes. In the left panel of Fig.\,\ref{fig:electron75LDF1} we also show the lateral distribution of the electric field in electron-induced showers.   In this case the differences between ZHAireS and CoREAS are $\lesssim3\%$ in the $-1.5\lesssim\alpha/\alpha_C\lesssim1.5$ range.

\subsubsection{Frequency spectrum}

In the right panels of Figs.\,\ref{fig:proton45LDF1}, \ref{fig:iron75LDF1} and \ref{fig:electron75LDF1} we show the average frequency spectrum of the $E_{y}$ component for CoREAS (blue dashed line) and ZHAireS (red dashed line), for observers on the ground at two distances to the shower core, near the \textit{Cherenkov ring} at off-axis angles $\alpha/\alpha_C\simeq1$ and farther from it at $\alpha/\alpha_C\simeq1.5$.  

The spectral behavior with frequency is similar in CoREAS and ZHAireS. For observers near the Cherenkov ring ($\alpha/\alpha_C \simeq 1$), the electric field rises linearly below $\sim 5\,\mathrm{MHz}$, flattens between 5 and 100 MHz, and decreases above 100 MHz. This can be understood in terms of the particle time distribution in the shower front as seen from the Cherenkov angle \cite{Ammerman-Yebra:2023rhr}. At low frequencies, all particles contribute nearly in phase, and the spectrum follows the single-particle track behavior \cite{Zas:1991jv}. The flattening from 5 MHz onward results from increasing decoherence due to time delays relative to a plane wavefront, caused by the curvature and width of the shower front. At high frequencies, decoherence increases further due to both longitudinal lag and transverse spread of particles, leading to a steeper spectral decline. For observers at $\alpha/\alpha_C \simeq 1.5$, the spectrum falls off at lower frequencies, reflecting larger time delays as they no longer observe the entire longitudinal development of the shower almost simultaneously \cite{Zas:1991jv,Ammerman-Yebra:2023rhr}.

In the 30 -- 80 MHz frequency range, relevant to many radio experiments~\cite{Schroder:2016hrv,Huege:2017khw}, ZHAireS generally predicts electric field amplitudes  $\lesssim 5\%$ smaller than CoREAS, with the trend reversing at lower frequencies. Above 300 MHz, differences grow progressively as the emission becomes largely incoherent and more sensitive to shower-to-shower fluctuations and to the thinning algorithms, which differ between both simulation packages\footnote{ZHAireS simulations show that variations in thinning parameters, especially the maximum allowed particle weight, have a stronger impact on the Fourier components of the electric field above 100 MHz than at lower frequencies.}. At these higher frequencies, discrepancies exceeding a few percent are expected, as seen in the right panels of Figs.\,\ref{fig:proton45LDF1}, \ref{fig:iron75LDF1}, and \ref{fig:electron75LDF1}, with ZHAireS systematically predicting lower emission than CoREAS. Comparable differences are also observed for off-axis observers located outside the Cherenkov cone, at $\alpha/\alpha_C \simeq 1.5$. 

\subsection{Energy in radio waves}
\label{S:Energy}

One of the most important observables in radio detection of air showers is the energy radiated in the form of radio waves (the \textit{radiation energy}) in a given frequency band. It has been experimentally demonstrated \cite{PierreAuger:2016vya} and also shown in simulations \cite{Glaser:2016qso, Gottowik:2017wio}, that the energy in radio waves serves as an accurate estimator of the electromagnetic energy in the shower that, at the same time, correlates with the energy of the primary particle. A comparison of the predicted energy in radio waves was performed in \cite{Gottowik:2017wio}, with CoREAS yielding $5.2\%$ higher radiation energy than ZHAireS in the 30 -- 80 MHz frequency band. As noted in the Introduction, this work uses a new version of ZHAireS that incorporates the same refractive index model of the atmosphere (based on the Gladstone-Dale law \cite{Gladstone-Dale}) as that used in CoREAS. This represents an improvement when compared to the exponential refractive index model employed in the ZHAireS version available at the time of \cite{Gottowik:2017wio}. Additionally, a newer version of CoREAS (v1.4) is used here, along with the same modern hadronic interaction models, tuned to LHC data, in both programs.

\begin{figure}[ht]
  \centering
  \begin{minipage}{0.48\linewidth}
    \centering
    \text{\small{CoREAS}}
    {\includegraphics[width=\linewidth]{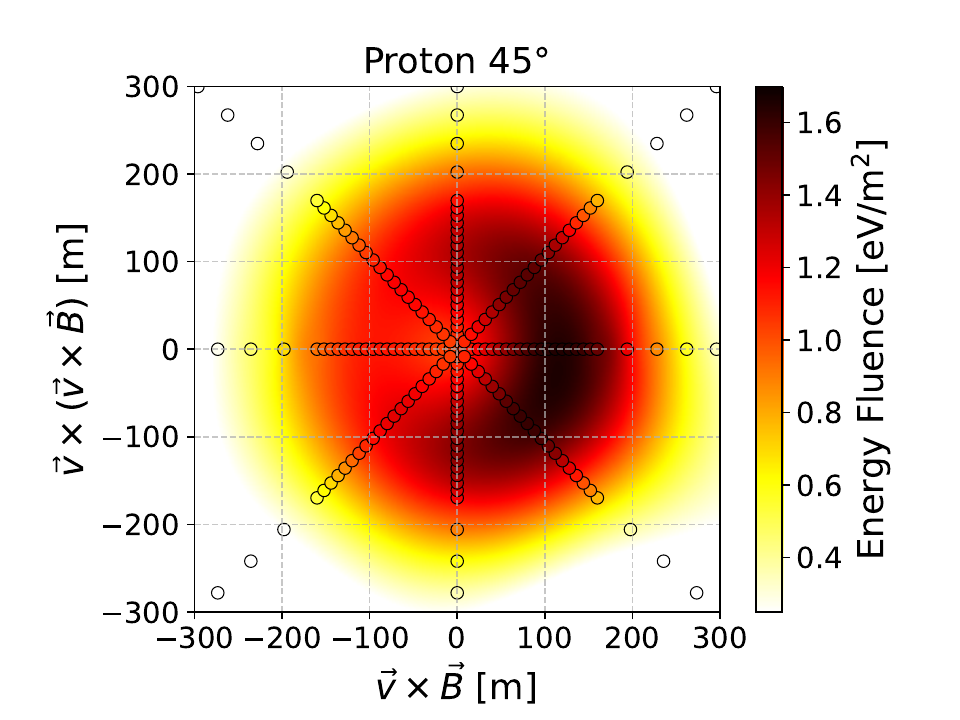}}
    {\includegraphics[width=\linewidth]{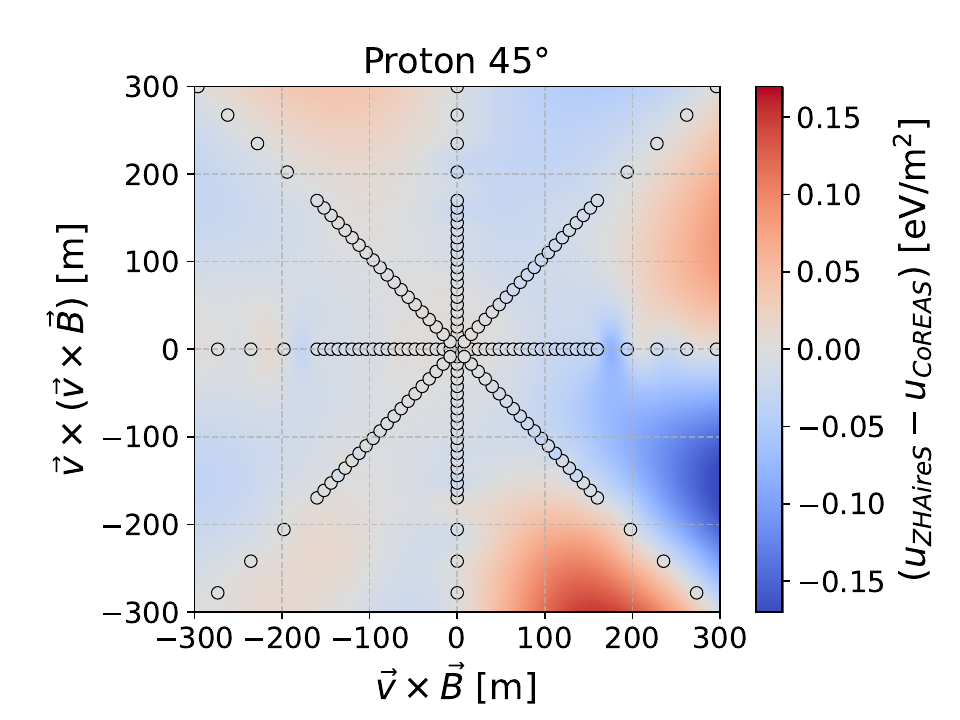}}
    \text{\small{Absolute difference}}
  \end{minipage}\quad
   \begin{minipage}{0.49\linewidth}
    \centering
    \text{\small{ZHAireS}}
    {\includegraphics[width=\linewidth]{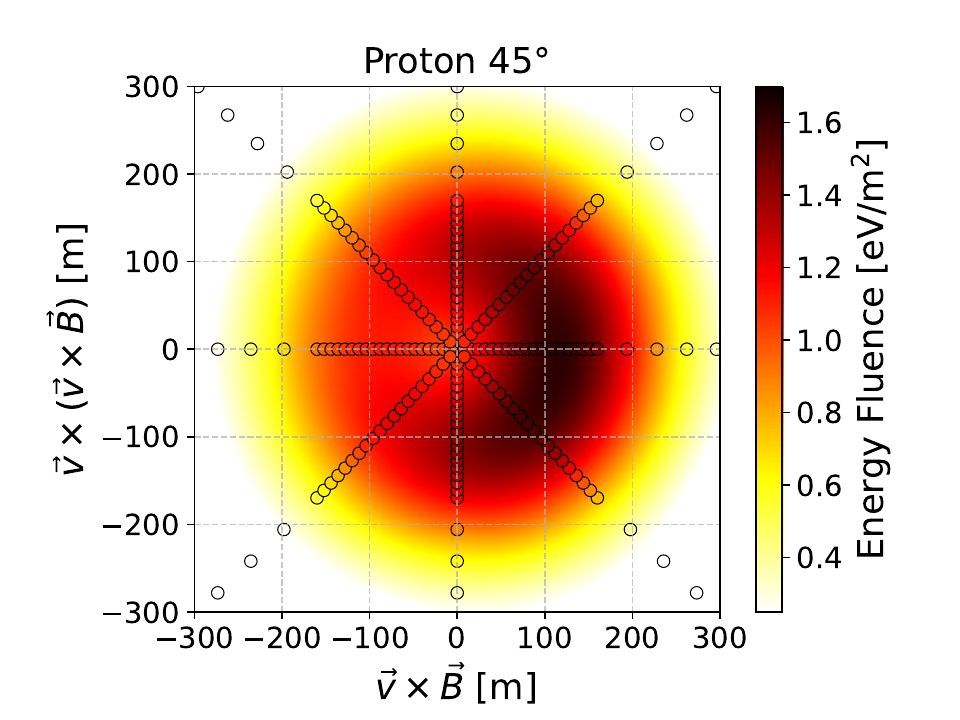}}
    {\includegraphics[width=\linewidth]{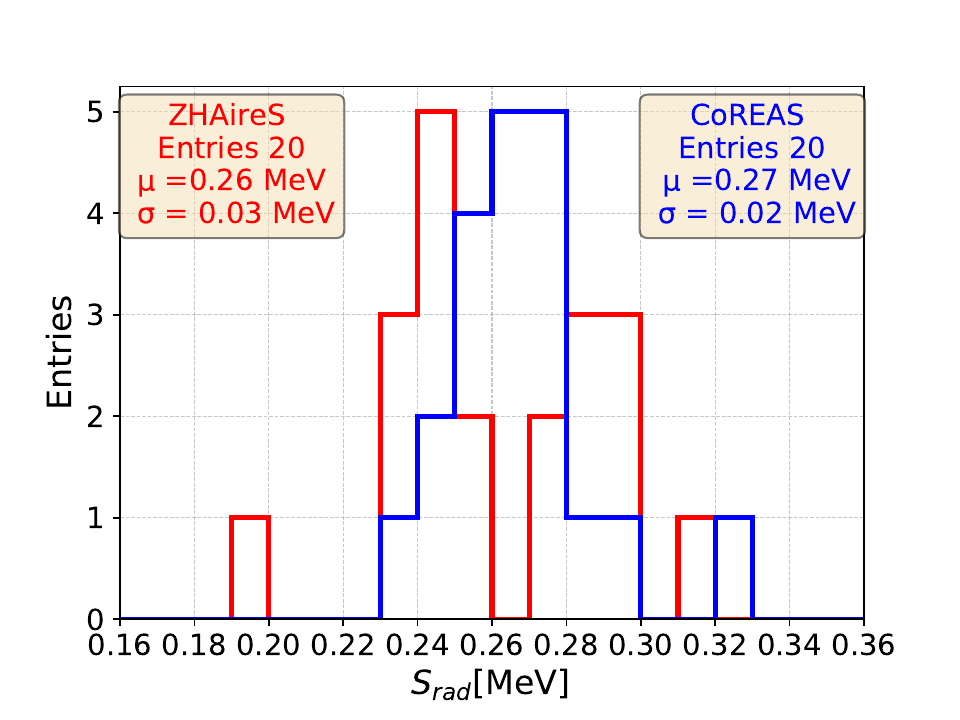}}
    \text{\small{Radiation energy distribution}}
  \end{minipage}\quad
   \caption{Top panels: Maps of the radio energy fluence emission (color scale) in the 30 -- 80 MHz band in the shower plane averaged over the 20 air showers simulated with CoREAS (top left) and ZHAireS (top right) for a proton primary of $E= 10^{17} \mathrm{eV}$ and $\theta=45^{\circ}$ with similar values of $X_\mathrm{max}$ within less than $1\,\mathrm{g/cm^2}$ of $X_\mathrm{max}=642.4\,\mathrm{g/cm^2}$. 
   The $x$ and $y$ axes correspond to distances along the $\hat{v}\times \vec{B}$ and $\hat{v}\times(\hat{v}\times \vec{B})$ directions respectively. Bottom left panel: Map of the absolute difference (ZHAireS-CoREAS) of the average energy fluence shown in the top panels. Bottom right panel: Histograms of the total radiated energy of the 20 showers simulated with ZHAireS (red) and CoREAS (blue). The average value ($\mu$) and dispersion around the mean ($\sigma$) are indicated.  
} 
   \label{fig:efluenceproton45}
\end{figure}

Following our methodology (Section\,\ref{S:Methodology}) of selecting simulated showers with values of $X_\mathrm{max}$ within $\lesssim 1\,\mathrm{g/cm^2}$ of a specified value, energy fluence maps in the 30 -- 80 MHz frequency band were generated for the same primaries, zenith angles and magnetic field configurations as in the previous plots. The energy fluence footprints on the ground plane, were projected onto the shower plane (perpendicular to the shower axis $\hat{v}$), and rotated so that the $x$-axis in the plots is parallel to $(\hat{v} \times \vec{B})$. In these maps, the energy fluence is evaluated at simulated antenna positions (represented by open circle markers) and interpolated between them for improved visibility.
A denser antenna configuration was chosen near the shower core to ensure accurate sampling, as the largest variations in the signal typically occur close to the core.

\begin{figure}[ht]
  \centering
  \begin{minipage}{0.48\linewidth}
    \centering
    \text{\small{CoREAS}}
      {\includegraphics[width=\linewidth]{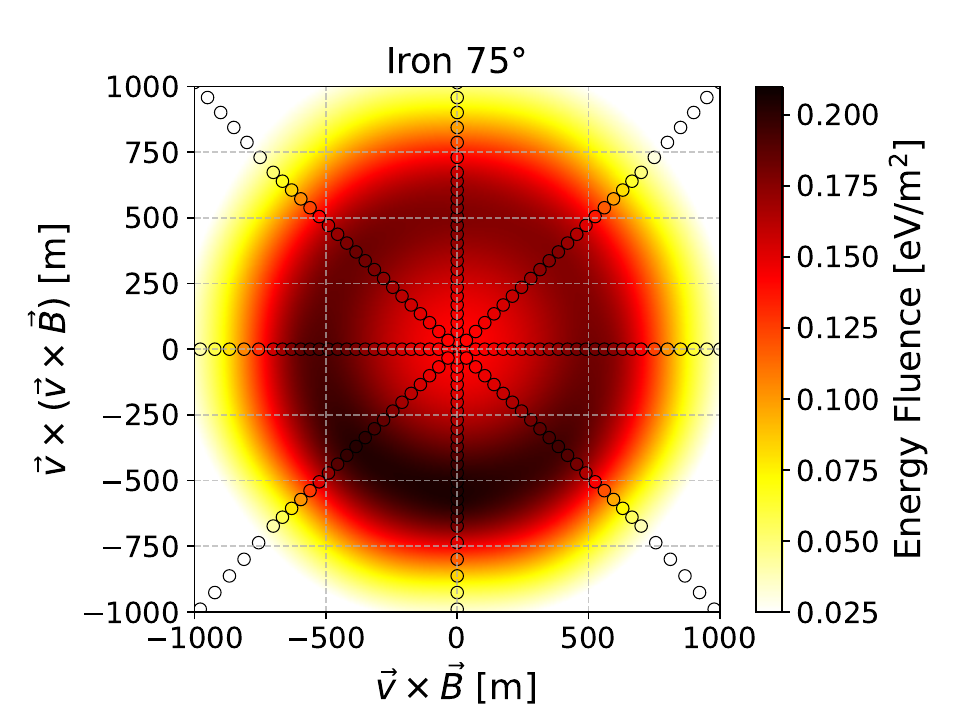}}
      {\includegraphics[width=\linewidth]{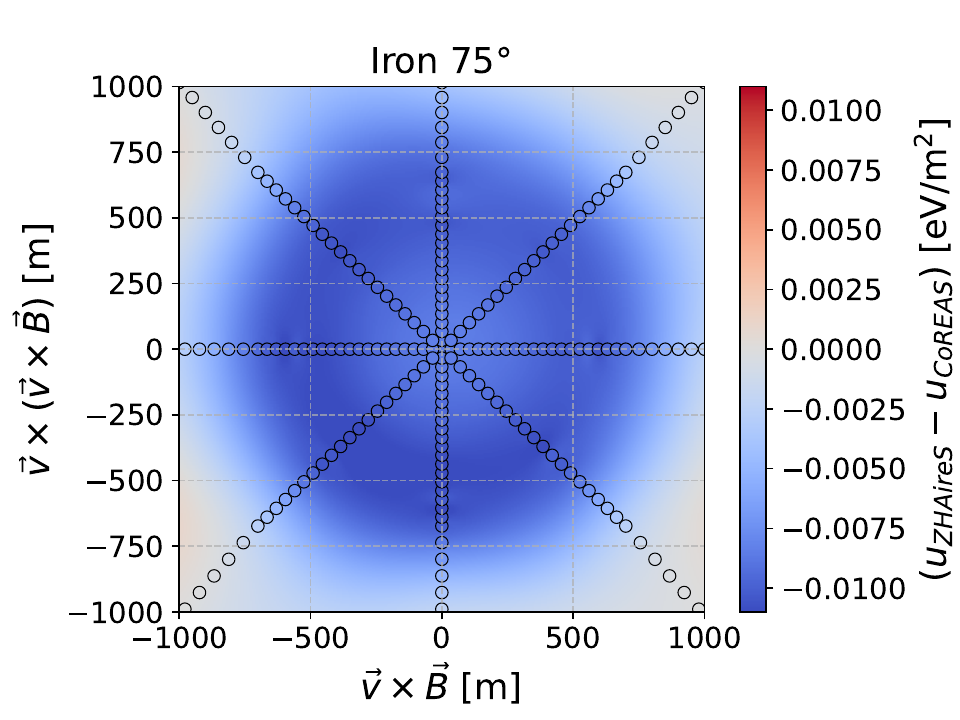}}
      \text{Absolute Difference}
   \end{minipage}\quad
   \begin{minipage}{0.49\linewidth}
   \centering
   \text{\small{ZHAireS}}
      {\includegraphics[width=\linewidth]{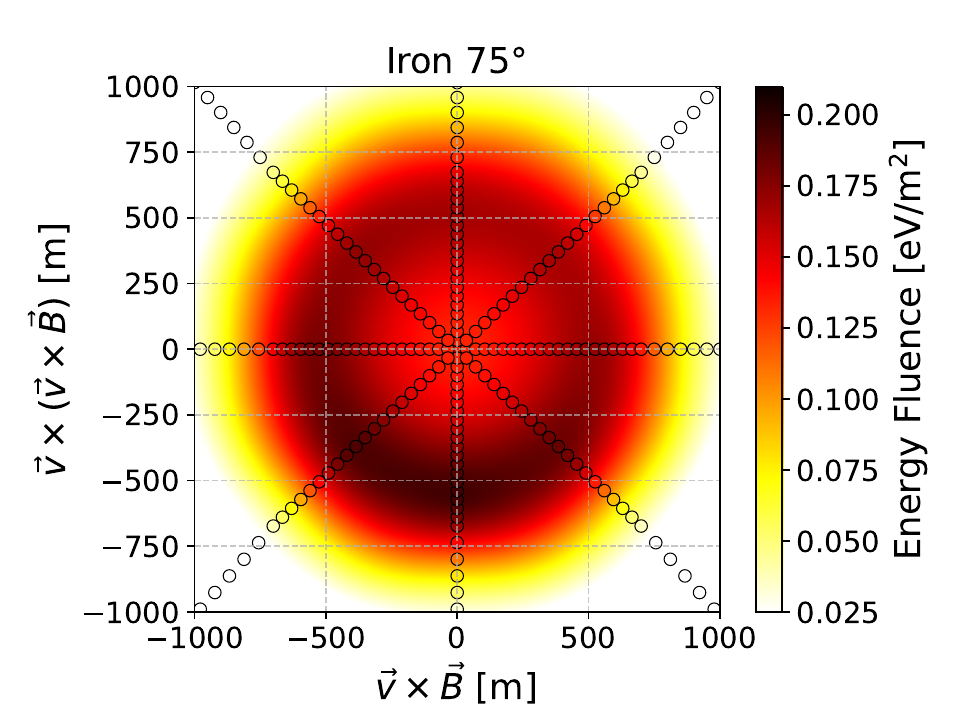}}
      {\includegraphics[width=\linewidth]{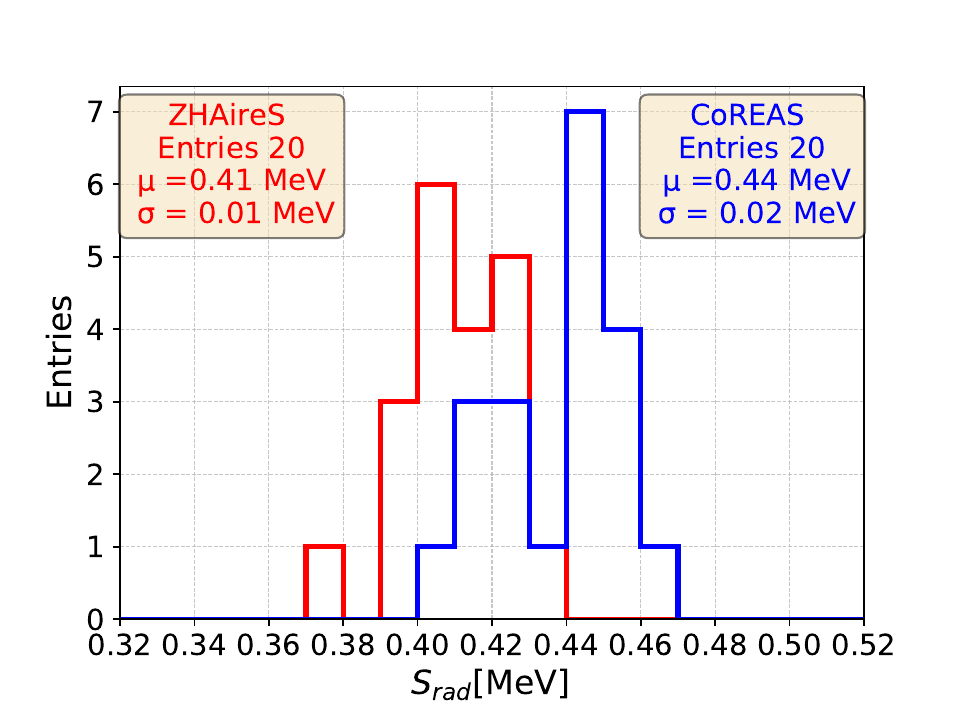}}
      \text{Radiation energy distribution}
  \end{minipage}\quad
   \caption{Same as Fig.\,\ref{fig:efluenceproton45} for iron primaries of $E= 10^{17} \mathrm{eV}$ and $\theta=75^{\circ}$ and a vertical field configuration with $|\vec{B}|=50\,\mu$T, perpendicular to the ground, with similar values of $X_\mathrm{max}$ within less than $1\,\mathrm{g/cm^2}$ of $X_\mathrm{max}=574.4\,\mathrm{g/cm^2}$.
   } 
   \label{fig:efluenceiron75}
\end{figure}

In the top panels of Fig.\,\ref{fig:efluenceproton45}, the energy fluence is shown for protons with zenith angle $\theta = 45^\circ$ for both CoREAS (top left) and ZHAireS (top right).  
The plot shows the fluence averaged over 20 showers, all having nearly the same $X_{\text{max}}\simeq 642.4~\mathrm{g/cm^2}$. As expected, a stronger signal is observed in the positive $(\hat{v} \times \vec{B})$ direction, where the  geomagnetic component adds to the Askaryan effect, when compared to the negative one where they subtract. 
At the bottom of Fig.\,\ref{fig:efluenceproton45}, the absolute difference between ZHAireS and CoREAS is presented, ranging between $\simeq 5\%$ and $\simeq-10\%$, the negative value implying that ZHAireS predicts a smaller energy fluence than CoREAS.  
In the right bottom panel of Fig.\,\ref{fig:efluenceproton45} we show the distribution of energy fluence for the 20 showers simulated with each code. The difference in the average value of the energy fluence is $\simeq -4\%$. The fluctuations of the radiated energy are similar in both programs corresponding to $\sigma\simeq 0.02\,$MeV. The mean radiation energy values obtained from both programs, CoREAS ($0.27$\,MeV) and ZHAireS ($0.26$\,MeV), are consistent within systematic uncertainties with the experimental measurement from the AERA detector at the Pierre Auger Observatory\,\cite{PierreAuger:2016vya}, which reports a value of approximately $0.34\pm 0.14$\,MeV. This comparison accounts for the expected scaling of the radiation energy with the primary energy ($E=10^{17}$\,eV), the magnetic field strength ($B=50\,\mu\mathrm{T}$), and the sine of the geomagnetic angle ($\sin\beta=0.7$) used in our simulations.

In Fig.\,\ref{fig:efluenceiron75}, the average energy fluence maps are presented for 20 iron showers of $E=10^{17}$ eV and $\theta=75^\circ$, all having nearly the same $X_{\text{max}}$. Due to the larger zenith angle and the corresponding larger distance from $X_{\text{max}}$ to ground, the signal spreads over a larger footprint on the shower plane when compared to those in Fig.\,\ref{fig:efluenceproton45} and the fluence is smaller. When integrating over the area to obtain the total radiation energy, the larger area compensates for the lower fluence values, resulting in total radiation energies of $0.44$\,MeV for CoREAS and $0.41$\,MeV for ZHAireS. These values are in agreement with the experimental result $\simeq 0.64\pm0.27$\,MeV \cite{PierreAuger:2016vya}, scaled for $E=10^{17}$\,eV, $B=50\,\mu\mathrm{T}$, and $\sin\beta=0.97$ of the showers simulated in this study.
The pattern of the fluence footprint also becomes more symmetrical than that at $\theta=45^\circ$, as for this geometry the magnitude of the Askaryan component is smaller relative to the geomagnetic one. This is in part due to the fact that the angle with respect to the geomagnetic field $\beta$ is closer to $90^\circ$, and in part due to the lower density of the atmosphere where the shower develops, allowing for stronger transverse currents~\cite{Glaser:2016qso}. As shown in the bottom-left panel of Fig.\,\ref{fig:efluenceiron75}, the absolute difference in the energy fluence between ZHAireS and CoREAS is within 10\% for all positions on the shower plane. The difference in the total average radiated energy is $\simeq -7\%$ with ZHAireS predicting less radiation energy than CoREAS.

\subsection{Depth of shower maximum}
\label{S:Xmax}

Another important observable in air showers is the depth of maximum shower development $X_{\mathrm{max}}$, which is accessible to radio detectors~\cite{Buitink:2016nkf, Tunka-Rex:2015zsa, PierreAuger:2023lkx,Corstanje:2025wbc}. For a given primary energy, the shape of the radio footprint on ground depends strongly on the primary particle type and can be used to determine the cosmic-ray mass composition on an statistical basis. A heavier primary nucleus will, on average, interact higher up in the atmosphere and produce a wider footprint on the ground than a lighter one. The particle type itself is not a direct observable, but $X_{\mathrm{max}}$, which depends on the particle type, can be related to the shape of the shower radio footprint. 

Several methods have been developed to reconstruct $X_{\text{max}}$ from radio signals \cite{Buitink:2014eqa,Tunka-Rex:2015zsa, CARVALHO201941}. 
The highest resolution has been achieved thus far by exploiting the lateral distribution of the radio footprint, fitting simulated air showers to measured air showers \cite{Buitink:2014eqa, Buitink:2016nkf}. This reconstruction method has been successfully applied to determine the depth of maximum with data collected at the AERA array of the Pierre Auger Observatory \cite{PierreAuger:2023lkx,PierreAuger:2023rgk}, achieving a resolution of better than $\simeq 15\,\mathrm{g/cm^2}$ at energies close to $10^{19}$ eV, as well as with data collected at Tunka-Rex \cite{Bezyazeekov:2018yjw}, LOFAR \cite{Corstanje:2021kik},  
and in simulation studies for the not-yet-operational SKA detector \cite{Corstanje:2025wbc}.

In this Section, for illustrative purposes, we apply a modified and largely simplified $X_{\text{max}}$ reconstruction method based on~\cite{Buitink:2014eqa,CARVALHO201941}. The method relies heavily on simulations of radio emission from air showers, and we use both CoREAS and ZHAireS to assess the reconstruction uncertainties arising from the choice of simulation code. It is important to stress that the aim here is not to validate the $X_{\text{max}}$ reconstruction method itself. Rather than that, we seek to estimate the uncertainty introduced by using different simulation programs in the reconstruction. Accordingly, uncertainties related to shower energy, geometry, and core position are not considered, as our focus is on establishing a lower bound on the intrinsic uncertainties due solely to the choice of simulation package. The absolute accuracy and applicability of other more involved methodologies under more realistic conditions have been studied  in~\cite{Buitink:2014eqa,CARVALHO201941} and demonstrated on experimental data~\cite{Buitink:2016nkf, Bezyazeekov:2018yjw, Corstanje:2021kik, PierreAuger:2023lkx,Corstanje:2025wbc}.

We simulated 250 proton- and 250 iron-induced showers with energy $E=10^{17}\,\text{eV}$, and zenith angle $\theta=45^\circ$, developing in the atmosphere under the influence of a magnetic field parallel to the ground with intensity $|\vec{B}| = 50 \, \mu\text{T}$, and pointing north. The simulations employed an array of 24 antennas arranged in a rectangular grid with 250\,m spacing.
The shower core was fixed at the center of the array, so that no antenna was located exactly at the origin.
Instead, the closest antennas were placed at a distance of approximately 125\,m from the core, with the grid symmetrically distributed around it.
This configuration was chosen to span roughly twice the typical diameter of the Cherenkov ring across the full range of $X_{\text{max}}$ values and zenith angles considered in our simulations, thereby ensuring that both the inner and outer regions of the Cherenkov ring were sampled depending on the shower development.
However, fixing the core to the array center does not capture the natural variability of core positions in real measurements.
While a denser grid would certainly provide higher spatial resolution and mitigate potential edge effects or overemphasis of individual antennas, the chosen spacing reflects array configurations such as AERA.

The reconstruction procedure starts by randomly selecting one of the 500 simulated showers to serve as the \textit{reference} to be reconstructed, treating it as mock data with a known depth of maximum, $X^\text{true}_{\text{max}}$. The remaining showers act as the \textit{reconstruction set}. This process is carried out for both simulation codes, by alternating the roles of the reference and reconstruction sets to evaluate cross-code consistency. Specifically, a single CoREAS-simulated shower is randomly chosen as the reference, and its $X_{\text{max}}$ is reconstructed using the other 499 CoREAS showers, or alternatively, the 500 ZHAireS showers. The same procedure is then repeated with a randomly selected ZHAireS shower, reconstructing it with either the remaining ZHAireS or CoREAS samples. This approach allows for both within-code and cross-code comparisons.

For each shower in the reconstruction set, we compute a quantity denoted $\Delta^2$, that quantifies the deviation between the radio signals of the reference shower and those of each of the showers in the reconstruction set. To account for possible differences in the overall amplitude of the electric field, we introduce a scaling factor $f_{r}$, which is adjusted independently for each simulated shower in the reconstruction set in order to minimize the deviation. $\Delta^2$ is defined as the quadratic sum of the differences between the energy fluence
of the electric field filtered between 30 and 80 MHz induced in the antennas of the reference shower,
${u}^{\,\text{ref}}_i$, and the scaled fluences $f_{r}^2 \cdot {u}_i$ 
at each antenna for each of the 499 showers in the reconstruction set~\cite{CARVALHO201941}:

\begin{equation}
\Delta^2 = 
\sum_{\text{i=antennas}} \left({u_i}^{\text{ref}} - f_{r}^2\,{u_i} \right)^2~~,
\label{eq:delta2}
\end{equation}
where the sum runs over all antennas in both the reference shower and in the showers of the reconstruction set. The energy fluence (in units of $\mathrm{eV/m^2}$) is defined as the sum over the time bins of the electric field trace of the total electric field squared $u=\epsilon_0 c \Delta t \sum_k \vert \vec{E}(t_k) \vert^2$, with $\epsilon_o$ the vacuum  permittivity, $c$ the speed of light, and $\Delta t$ the size of the time bin \cite{PierreAuger:2023rgk}. The scaling factor $f_{r}$ is optimized for each shower in the reconstruction set to yield the minimum possible $\Delta^2$.  A value of $\Delta^2$ is obtained for each of the 499 simulated showers (excluding the reference one).  
 We identify the shower in the reconstruction set that yields the minimum $\Delta^2$ and take its $X_{\text{max}}$ as the reconstructed value, denoted $X^\text{rec}_{\text{max}}$. This procedure constitutes a largely simplified version of the $X_{\text{max}}$ reconstruction procedure introduced in~\cite{Buitink:2014eqa, CARVALHO201941}. 

To ensure an unbiased comparison, outliers in the $X_{\text{max}}$ distribution are excluded from the pool of showers randomly chosen to be reconstructed. This step prevents extreme cases from skewing the results and focuses the analysis on the bulk of the shower $X_\text{max}$ distribution. The simplicity of the method is intentional, ensuring that the comparison is not strongly affected by other sources of uncertainty such as model-dependent optimizations, fitting functions, or detector effects, and allowing for a more straightforward comparison between the CoREAS and ZHAireS simulation codes to isolate their impact on reconstruction.

\begin{figure}[htbp]
  \centering
  \begin{minipage}{0.48\linewidth}
    \centering
      \text{\footnotesize{ZHAireS proton, reconstructed with CoREAS}}
      {\includegraphics[width=\linewidth]{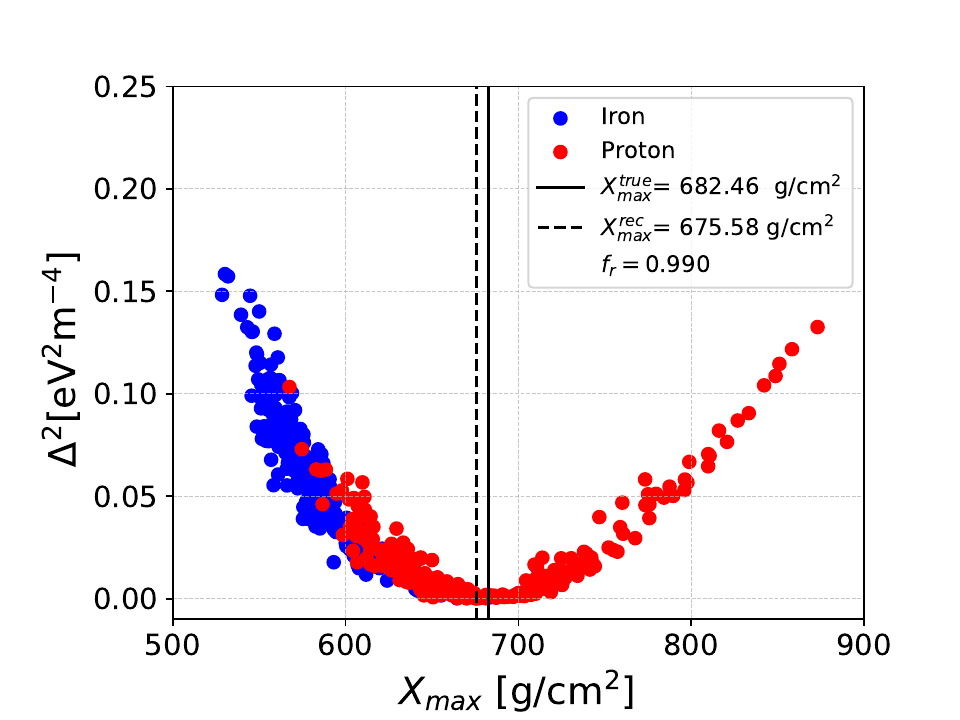}}
      \text{\footnotesize{CoREAS iron, reconstructed with CoREAS}}
      {\includegraphics[width=\linewidth]{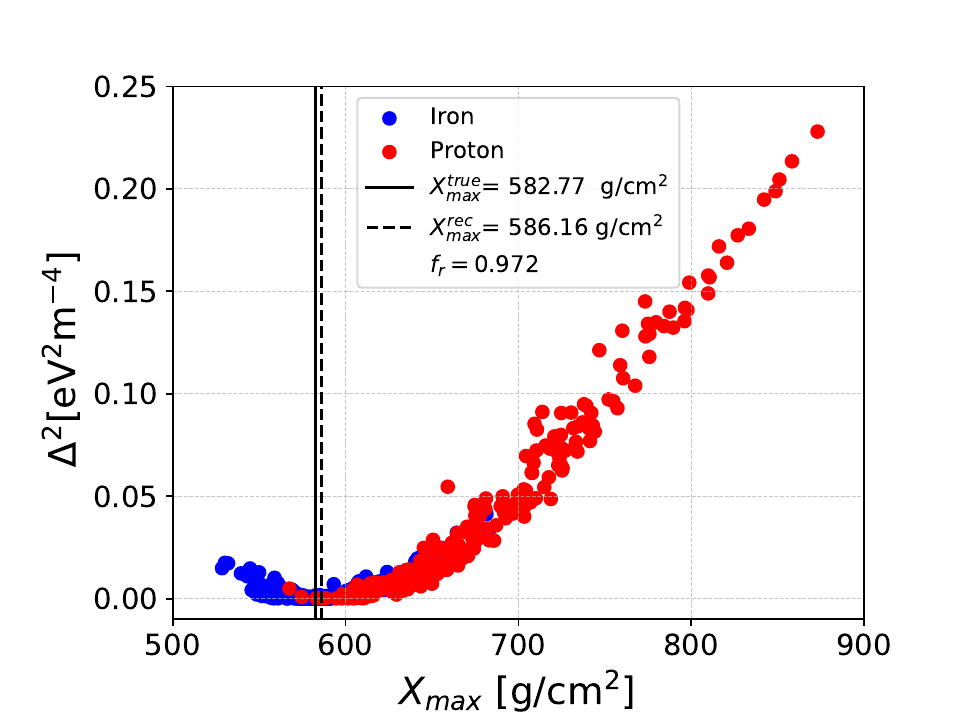}}
  \end{minipage}
  \begin{minipage}{0.48\linewidth}
    \centering
      \text{\footnotesize{ZHAireS proton, reconstructed with ZHAireS}}
      {\includegraphics[width=\linewidth]{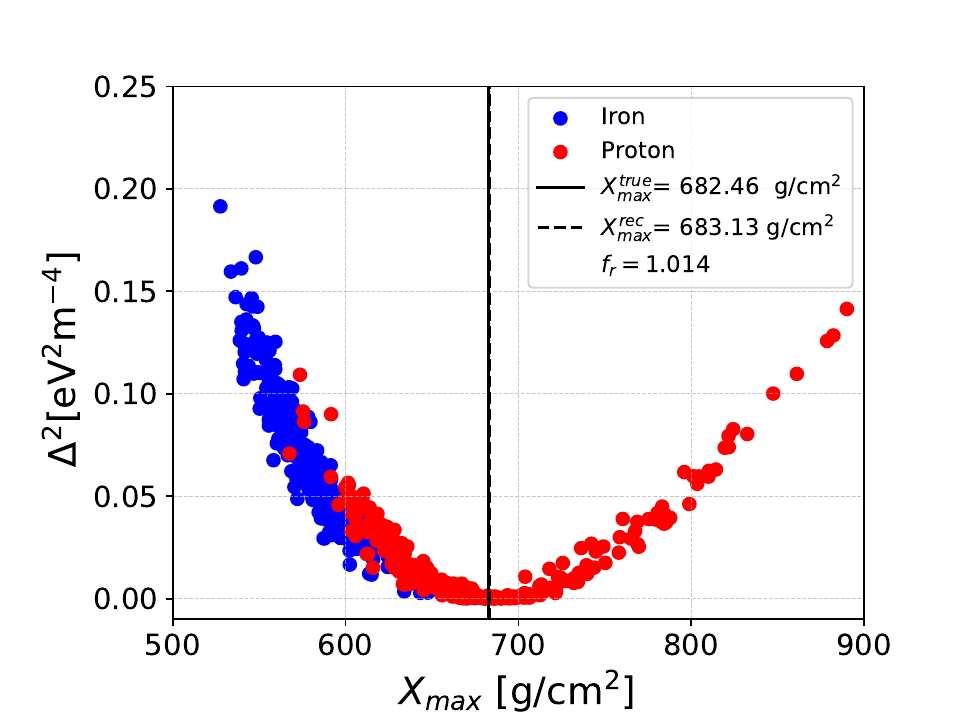}}      
      \text{\footnotesize{CoREAS iron, reconstructed with ZHAireS}}
      {\includegraphics[width=\linewidth]{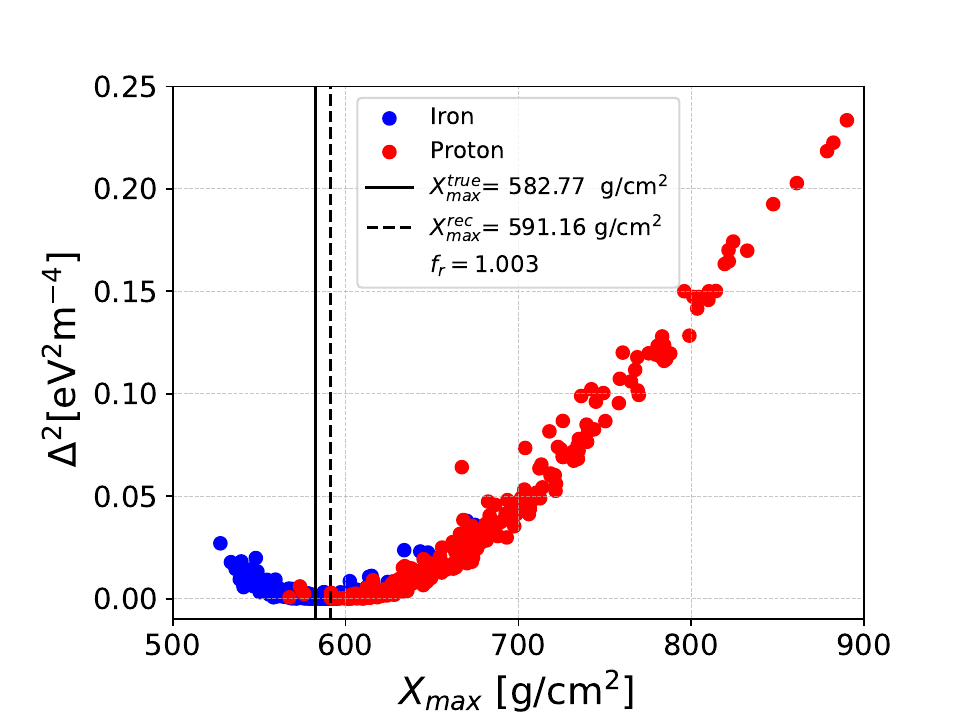}}
  \end{minipage}  
\caption{Four examples of the methodology used to reconstruct $X_{\mathrm{max}}$. For all cases, $E=10^{17},\mathrm{eV}$ and $\theta = 45^\circ$, the value of $\Delta^2$ in Eq.(\ref{eq:delta2}) (given in $ \text{eV}^{2}\text{m}^{-4}$ ) is plotted as a function of simulated $X_{\mathrm{max}}$ for the CoREAS and ZHAireS showers in the reconstruction sets, excluding the shower to be reconstructed with depth of maximum indicated with a black solid vertical line. The reconstructed $X^\text{rec}_\text{max}$ is indicated with a dashed black solid line. Red and blue points represent proton and iron simulations respectively. Top panels: the shower to be reconstructed is induced by a proton simulated with ZHAireS and reconstructed with the 500 CoREAS template showers (top left) or the remaining 499 ZHAireS template showers (top right). Similarly, in the bottom panels an iron reference shower is simulated with CoREAS and reconstructed using the remaining 499 CoREAS (bottom left) or with the 500 ZHAireS template showers (bottom right). See text for further details.}
  \label{fig:xmaxrecon1}
  \end{figure}

\begin{figure}[htbp]
  \centering
  \begin{minipage}{0.48\linewidth}
    \centering
      \text{\footnotesize{ZHAireS proton, reconstructed with CoREAS}}
      {\includegraphics[width=\linewidth]{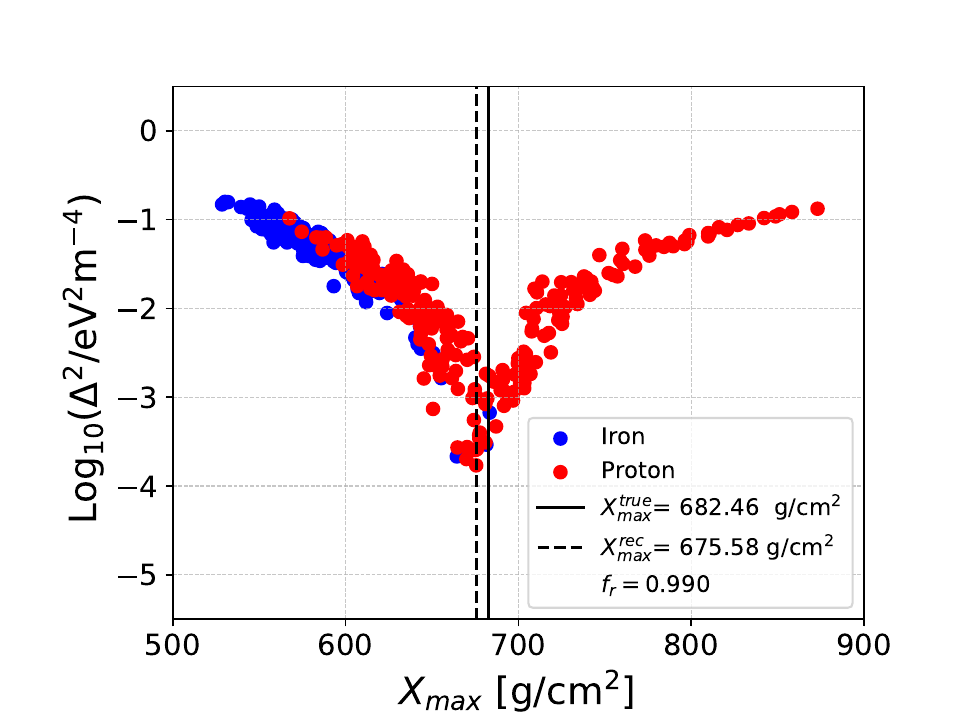}}
      \text{\footnotesize{CoREAS iron, reconstructed with CoREAS}}
      {\includegraphics[width=\linewidth]{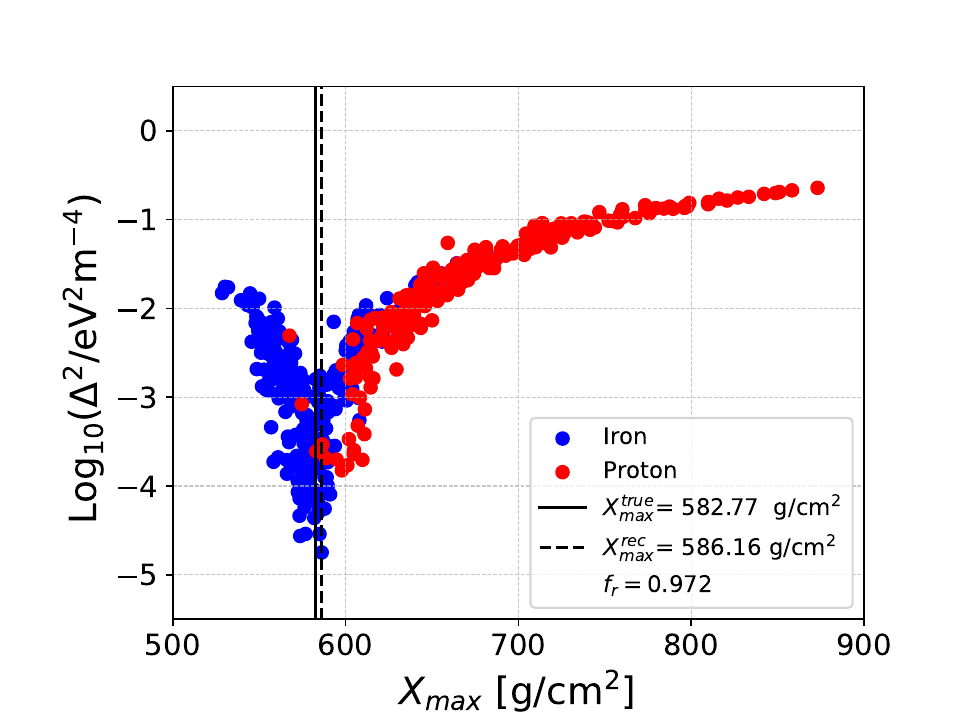}}
  \end{minipage}
  \begin{minipage}{0.48\linewidth}
    \centering
      \text{\footnotesize{ZHAireS proton, reconstructed with ZHAireS}}
      {\includegraphics[width=\linewidth]{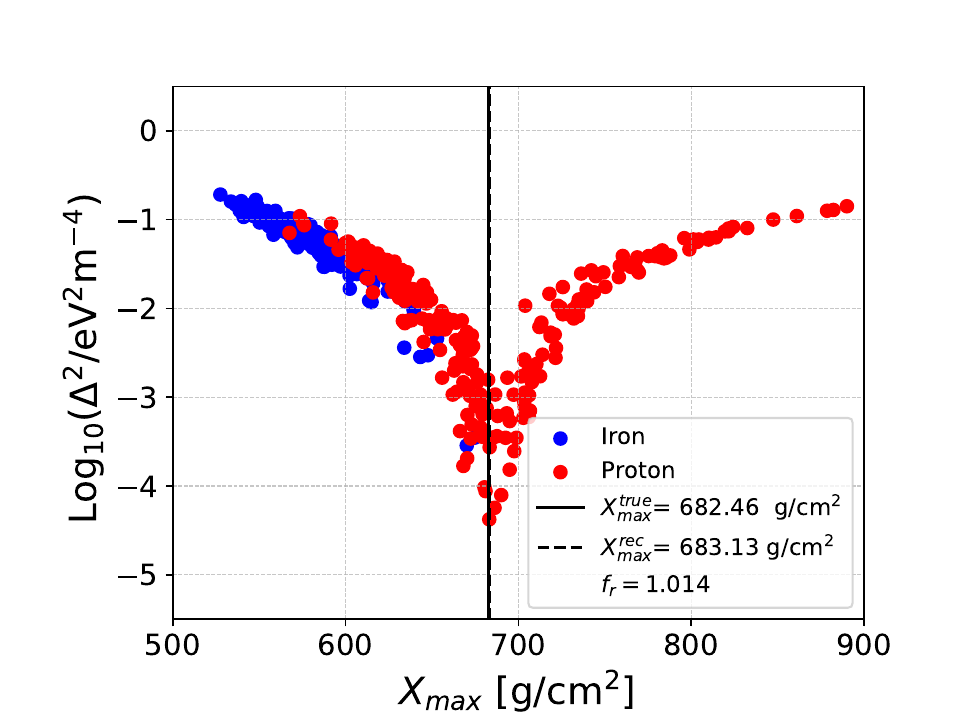}}      
      \text{\footnotesize{CoREAS iron, reconstructed with ZHAireS}}
      {\includegraphics[width=\linewidth]{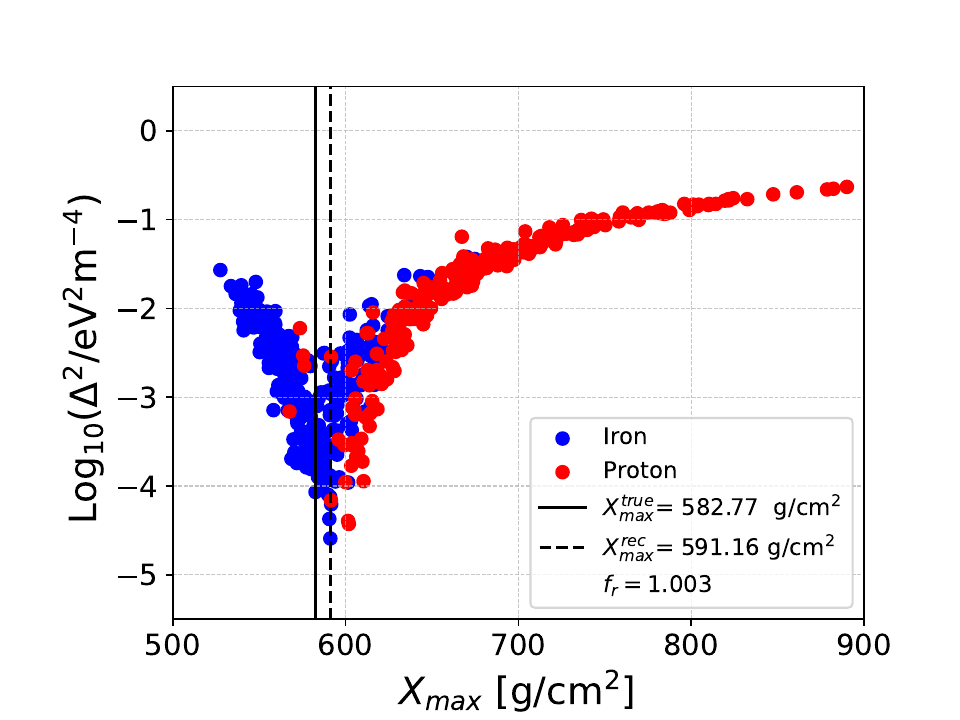}}
  \end{minipage}  
\caption{Same as Fig.\,\ref{fig:xmaxrecon1} with a logarithmic scale in the y-axis.}
  \label{fig:xmaxreconlog}
  \end{figure}

Figs.\,\ref{fig:xmaxrecon1} and \ref{fig:xmaxreconlog} present four illustrative examples of the reconstruction method. In the top panels, a proton shower simulated with ZHAireS is used as the reference, and its $X_{\text{max}}$ is reconstructed using either the 500 CoREAS (left) or the remaining 499 ZHAireS (right) showers as the reconstruction set. The bottom panels show the reverse case: an iron shower simulated with CoREAS is reconstructed using either the 499 CoREAS (left) or the 500 ZHAireS (right) templates. In all panels, $\Delta^2$ is plotted as a function of the $X_{\text{max}}$ of each shower in the reconstruction set, excluding the reference shower. Red and blue points indicate proton and iron primaries, respectively.
These examples illustrate how the method identifies the best-matching shower, regardless of whether the reconstruction is performed within the same code or across codes. This is further emphasized in Fig.~\ref{fig:xmaxreconlog} where a logarithmic scale is used to display $\Delta^2$. In all four cases, the reconstructed $X_{\text{max}}$ differs from the true value by less than $\simeq 10\,\mathrm{g/cm^2}$, showing that even this simplified approach yields a reasonably accurate estimate of the shower maximum.

To compare the reconstruction bias of $X_\text{max}$ both within and across simulation codes, the procedure was repeated using 450 randomly selected proton and iron showers from each of the CoREAS and ZHAireS samples of 500 showers (excluding outliers). Each selected shower was treated as a reference and reconstructed using either the CoREAS or ZHAireS templates, allowing again for both within-code and cross-code comparisons. For each case, the difference between the true and reconstructed depth, $X^\text{true}_{\text{max}} - X^\text{rec}_{\text{max}}$, was obtained.

\begin{figure}[htbp]
 \centering
  \begin{minipage}{0.48\linewidth}
    \centering
      {\includegraphics[width=\linewidth]{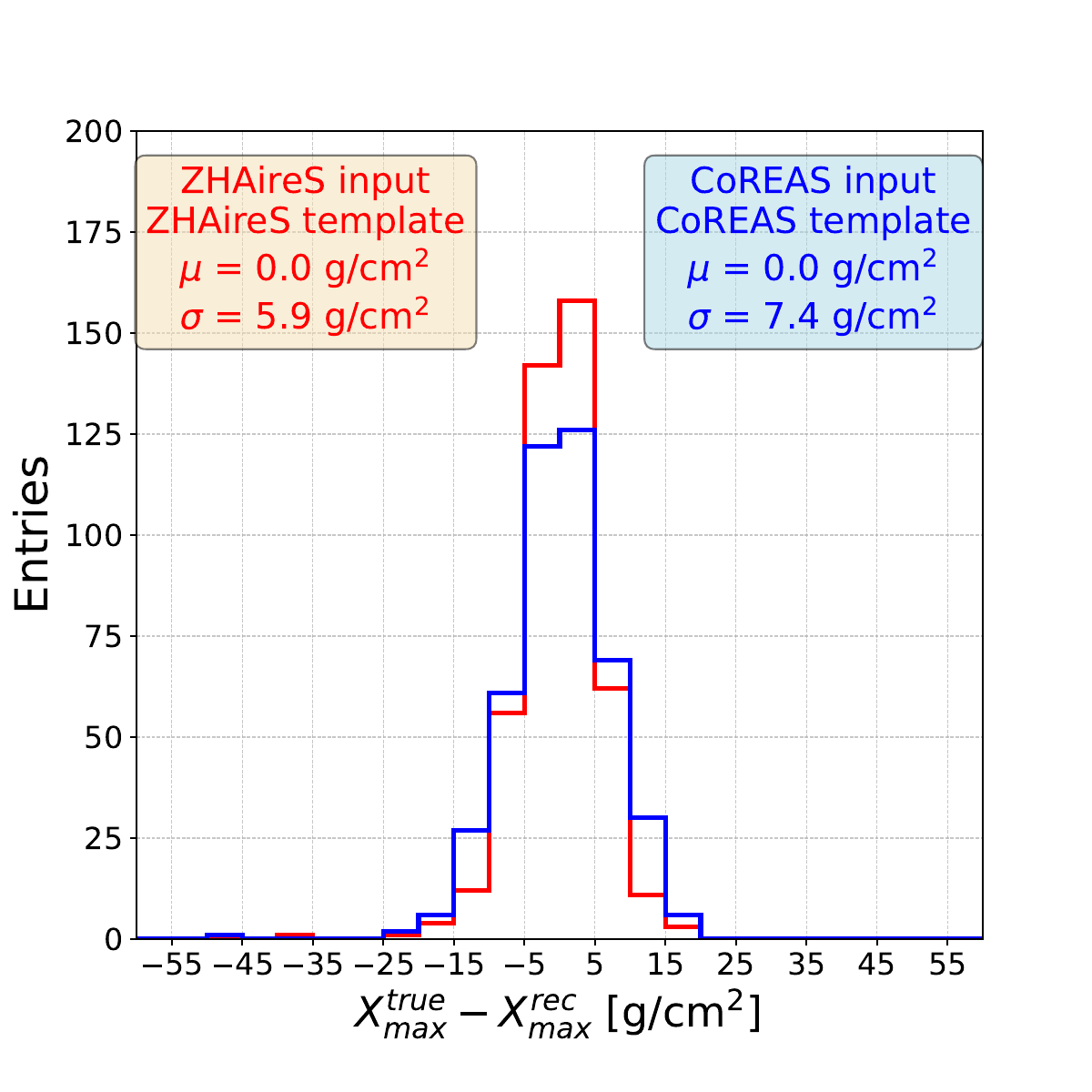}}
      {\includegraphics[width=\linewidth]{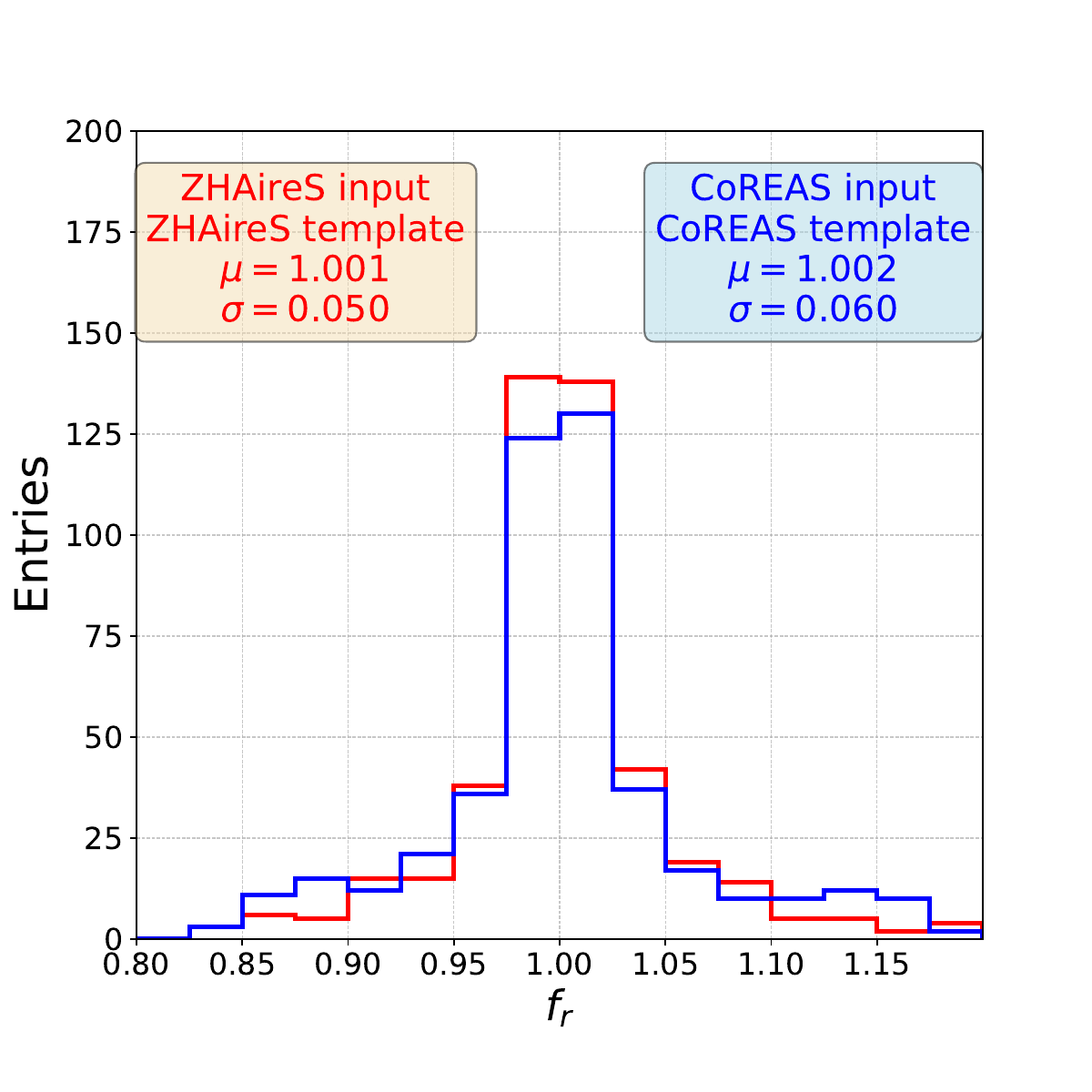}}
  \end{minipage}
  \begin{minipage}{0.48\linewidth}
    \centering
      {\includegraphics[width=\linewidth]{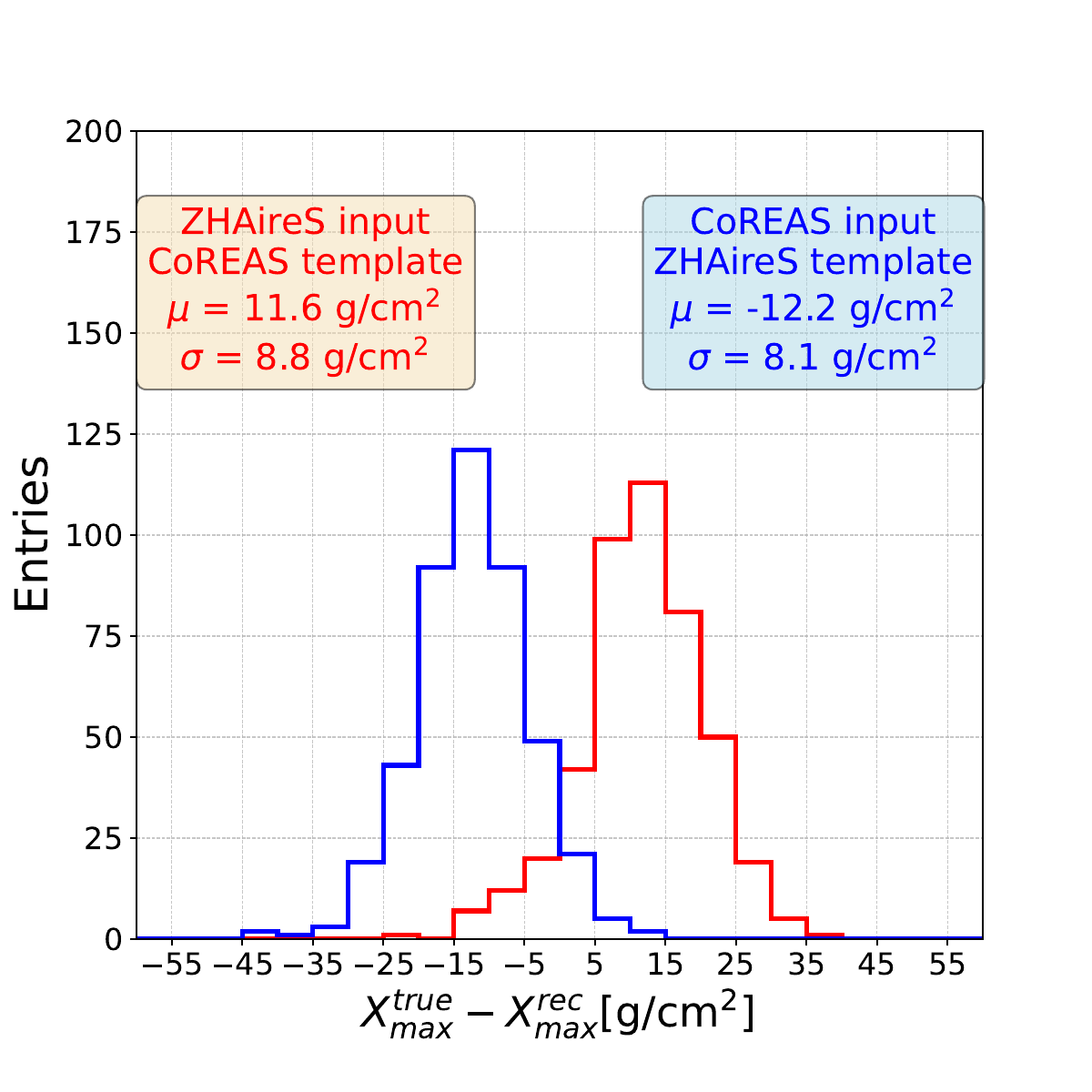}} 
      {\includegraphics[width=\linewidth]{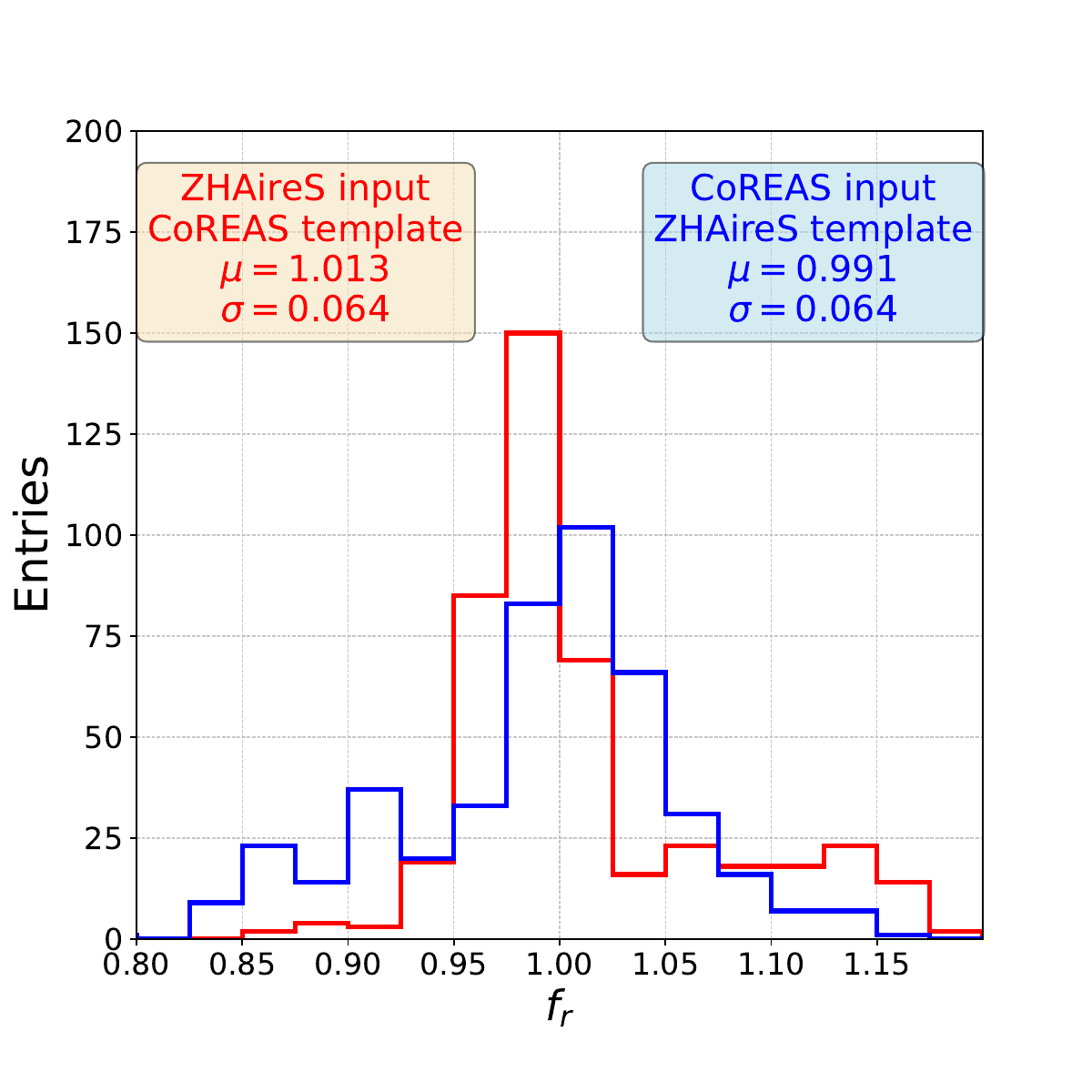}}
  \end{minipage} 
\caption{Top panels: Distribution of $[X_{\text{max}}^{\text{true}} - X_{\text{max}}^{\text{rec}}]$ $(\text{g/cm}^2)$ for 450 reference proton and iron simulated showers with $E=10^{17}\,\mathrm{eV}$ and zenith angle $\theta=45^\circ$. In blue (red) we show the results for the CoREAS (ZHAireS) simulations. The mean (indicating the bias) and RMS of each distribution are given. Left panel: Input showers to be reconstructed and template showers come from the same simulation code (ZHAireS in red, CoREAS in blue). Right: Input showers and template showers are simulated with different codes, labeled as \textit{ZHAireS input CoREAS template} for input showers simulated with ZHAireS and reconstructed with CoREAS, and as \textit{CoREAS input ZHAireS template} for input showers simulated with CoREAS and reconstructed with ZHAireS. 
Bottom panels: Corresponding distributions of the scaling factor $f_r$ introduced in Eq.\,(\ref{eq:delta2}).}
\label{fig:xmaxrecon2}
\end{figure}

The resulting distributions of these differences are shown in Fig.\,\ref{fig:xmaxrecon2}. When reconstruction is performed within the same simulation code (top-left panel), the mean value shows no significant bias, and the $\sigma$ values range from $5.9$ to $7.4\,\mathrm{g/cm^2}$. This indicates that using either code consistently for reconstructing $X_\mathrm{max}$ with this simplified method introduces a negligible uncertainty.

In contrast, when the reference showers are simulated with one code but reconstructed using templates 
from the other (top-right panel), the mean values shift to $\mu \simeq +11.6 \pm 0.4$ and $-12.3 \pm 0.6 \,\mathrm{g/cm^2}$, 
while the $\sigma$ values increase to $8.8$ and $8.1\,\mathrm{g/cm^2}$, respectively.

The two lower panels of Fig.,\ref{fig:xmaxrecon2} show the distributions of the scaling factor $f_r$ corresponding to the results in the upper panels. Since $f_r$ is defined as the ratio of the reference to the reconstructed shower energy, its values are expected to cluster around unity, as indeed seen in the histograms. When the same code is used for both simulation and reconstruction, the mean value is $\langle f_r\rangle \simeq 1.00$ (bottom left panel). In contrast, when different codes are used, $\langle f_r\rangle$ shifts below unity for showers simulated with CoREAS and reconstructed with ZHAireS, and above unity in the reverse case (bottom right panel), consistent with the generally larger field strengths predicted by CoREAS.

\section{Summary and conclusions}
\label{S:Summary and conclusions}

This work presents a comprehensive comparison between the two leading Monte Carlo simulation codes for radio emission from extensive air showers: CoREAS and ZHAireS. Both programs were tested under similar conditions, using identical primary particles, geomagnetic configurations, almost identical depth of shower maximum to reduce shower-to-shower fluctuations, and realistic atmospheric profiles derived from the Gladstone-Dale law. To ensure reproducibility, the full set of input directives and simulation settings for both codes is provided.

The results show a high level of consistency in the predicted radio signals. The lateral distribution of the electric field at 50 MHz agrees within $\lesssim 5\%$ around the Cherenkov angle. In the 30 -- 80 MHz range, which is relevant for many radio experiments~\cite{Schroder:2016hrv,Huege:2017khw}, ZHAireS typically predicts field amplitudes up to $\lesssim 5\%$ smaller than CoREAS, while at lower frequencies this trend reverses. At frequencies above 300 MHz, differences become more pronounced due to the loss of coherence and the increased sensitivity to shower-to-shower fluctuations and to the thinning algorithms, which differ between the two codes. In general, ZHAireS yields systematically smaller field amplitudes than CoREAS.

A strong linear correlation between the electric field magnitude and the electromagnetic energy content of the shower was observed in both codes, with higher electromagnetic fractions in CoREAS resulting in stronger radio signals.

Energy fluence maps for both proton and iron primaries also show very good agreement, with average differences remaining below $\lesssim 10\%$, and comparable shower-to-shower fluctuations. Furthermore, the mean radiation energy predicted by both codes is consistent within systematic uncertainties with the experimental measurements from the AERA detector at the Pierre Auger Observatory.

In this work, we have also investigated the impact of the simulation code choice on the reconstruction of the depth of shower maximum, $X_{\text{max}}$, by comparing \textsc{CoREAS} and \textsc{ZHAireS} using a simplified reconstruction method. We find that the uncertainty introduced solely by employing different codes amounts to $\lesssim 12$~g/cm$^2$. This indicates that non-negligible differences exist between the two simulation codes, and highlights the need for further studies to investigate their origin and to assess their impact on $X_{\text{max}}$ inference in cosmic-ray composition analyses.  

Identifying the origin of these differences is highly challenging, as contributions from shower evolution and radio-emission modeling are difficult to disentangle. \textsc{CoREAS} and \textsc{ZHAireS} rely on different approximations for energy-loss modeling ($dE/dX$), on different low-energy hadronic interactions below a few hundred GeV, and also implement inherently different thinning algorithms despite efforts to align their performance through parameter tuning. Fully separating these effects would require substantial modifications to both codes, which are beyond the scope of the present work. 

Nevertheless, the fact that the remaining discrepancies amount to only about 5\% in the electric field, suggests that the overall level of agreement between CoREAS and ZHAireS is already remarkable. A dedicated effort to disentangle the contributions from shower physics and radio-emission approximations would be valuable to improve our understanding and to strengthen the reliability of radio-based cosmic-ray analyses. 

Finally, our study indicates that applying both CoREAS and ZHAireS in data analyses may be advisable for cross-validation, at least until these discrepancies are fully understood.
\section{Acknowledgments}

This work has received financial support from
Ministerio de Ciencia, Innovaci\'on y Universidades/Agencia Estatal de Investigaci\'on MICIU/AEI /10.13039/501100011033, Spain
(PID2022-140510NB-I00, PCI2023-145952-2, CNS2024-154676, and Mar\'\i a de Maeztu grant CEX2023-001318-M);
Xunta de Galicia, Spain (CIGUS Network of Research Centers,  
Consolidaci\'on ED431C-2025/11, and 2022 ED431F-2022/15);
Feder Funds;
and European Union ERDF.
The authors
acknowledge the Subsecretaría de Ciencia y Tecnolog\'\i a, Argentina, and
Sistema Nacional de Computaci\'\o n de Alto Desempe\~o de la Rep\'ublica
Argentina for the use of its Clementina XXI supercomputer. 
P.M.H thanks ``Subsidio para viajes y estancias de la Facultad de
Ingenier\'\i a - UNLP", and thanks the Astroparticle Physics group at IGFAE, Univ. of Santiago de Compostela for their hospitality. We thank M. Seco (IGFAE-USC) for computing resources.

\bibliography{bibliography.bib}

\end{document}